\documentclass{aa}

\usepackage[varg]{txfonts}
\usepackage{epsfig,graphicx,natbib,url,twoopt}
\usepackage[varg]{txfonts}
\usepackage{hyperref}          
\hypersetup{
 colorlinks=true,  
 urlcolor=blue,    
 linkcolor=red     
}

\def\kms{\hbox{km$\;$s$^{-1}$}}
\def\deg{\hbox{$^\circ$}}      

\def\Halpha{\mbox{H\hspace{0.1ex}$\alpha$}}

\def\CaIR{\ion{Ca}{ii}~854.2\,nm}
\def\CaK{\ion{Ca}{ii}~K}
\def\CaH{\ion{Ca}{ii}~H}
 
\def\Hinode{{Hinode}}
\def\IRIS{{IRIS}}
\def\Lyalpha{\mbox{Ly\hspace{0.1ex}$\alpha$}}

\begin{document}

\title{Penumbral microjets at high spatial and temporal resolution}

\author{Luc H. M. Rouppe van der Voort\inst{1,2} 
\and
Ainar Drews\inst{1,2} 
}
\authorrunning{Rouppe van der Voort \& Drews}

\institute{Institute of Theoretical Astrophysics,
  University of Oslo, %
  P.O. Box 1029 Blindern, N-0315 Oslo, Norway
\and
 Rosseland Centre for Solar Physics,
  University of Oslo, %
  P.O. Box 1029 Blindern, N-0315 Oslo, Norway}

\date{Received 22 February 2019 / Accepted 7 May 2019}


\abstract
{
Sunspot observations in chromospheric spectral lines have revealed short-lived linear bright transients that are commonly referred to as penumbral microjets (PMJs).
Details on the origin and physical nature of PMJs are to a large extend still unknown.
}
{
We aim to characterize the dynamical nature of PMJs to provide guidance for future modeling efforts.
}
{
We analyzed high spatial (0\farcs1) and temporal resolution (1~s) \ion{Ca}{ii}~H filtergram (0.1~nm bandwidth) observations of a sunspot that were obtained on two consecutive days with the Swedish 1-m Solar Telescope.
}
{
We find that PMJs appear to be the rapid brightening of an already existing (faint) fibril. The rapid brightening is the fast increase (typically less than 10~s) in intensity over significant length (several hundreds of kilometers) of the existing fibril. For most PMJs, no clear root or source from where the brightening appears to originate can be identified. After the fast onset, about half of the PMJs have tops that move with an apparent velocity of between 5 and 14~\kms, most of them upward. No significant motion of the top is observed in the other PMJs. About one-third of the PMJs split into two parallel and coevolving linear features during the later phases of their lifetimes. 
}
{
We conclude that mass flows can play only a limited role in the onset phase of PMJs. It is more likely that we see the effect of a fast heating front. 
}

\keywords{Sun: activity -- Sun: atmosphere -- Sun: magnetic fields}

\maketitle

\section{Introduction}
\label{sec:introduction}

Observations in the chromospheric \ion{Ca}{ii} lines show dynamic, linear transients in sunspot penumbrae.
These so-called penumbral microjets (PMJs) were discovered
\citep{2007Sci...318.1594K} 
in \CaH\ time series from the broadband filtergraph (BFI) at the Solar Optical Telescope
\citep[SOT,][]{2008SoPh..249..167T} 
of the \Hinode\ satellite
\citep{2007SoPh..243....3K}. 
In the \Hinode\ observations, PMJs appear as linear brightenings, typically $\sim$400~km wide and 1000--4000~km long, and they have lifetimes shorter than 1~min. 
The increase in brightness is on the order of 10--20\% as compared to the penumbral background, and the PMJs stand out against the nearly horizontal photospheric penumbral filaments with elevation angles of between 20$\deg$ to 60$\deg$.
Probably the most striking property of PMJs is their rapid appearance: \citet{2007Sci...318.1594K} reported an apparent rise velocity faster than 100~\kms\ starting from the root of the microjets during the initial phase of their evolution. %
\citet{2016ApJ...816...92T} 
argued that microjets in the inner penumbra and larger jets in the outer penumbra should be distinguished. They roughly estimated a speed of 250~\kms\ for one such large penumbral jet.
This is much faster than the chromospheric sound speed, which is on the order of 10~\kms. 

Magnetic reconnection is the prime candidate process for driving PMJs: with strong magnetic fields with highly variable inclination angles over short spatial scales and considerable dynamic forcing from convective flows, the
``uncombed'' magnetic field topology of the sunspot penumbra is an environment where magnetic reconnection is likely to occur 
\citep[for reviews on the sunspot magnetic structure with strong-field vertical spines and horizontal filaments, see, e.g., ][]{2011LRSP....8....4B, 
2017arXiv171207174T}. 
\citet{2010A&A...524A..20K} found support for this scenario in small patches of photospheric downflows that are associated with chromospheric brightenings: the downflow may be the reconnection outflow, while the opposite-direction outflow may result in an associated PMJ.

Whether the fast apparent rise velocity of PMJs is a true mass flow or the result from a fast propagating wave or heating front cannot be determined from imaging data alone. 
Spectroscopic observations of the \CaIR\ line show a characteristic PMJ spectral profile with peaks in the wings at about $\pm10$~\kms\ 
\citep{2013ApJ...779..143R, 
2015ApJ...811L..33V, 
2017A&A...602A..80D}. 
The extent of these peaks is rarely found to be beyond $\pm20$~\kms\ and certainly never toward the extreme velocities of the apparent PMJ rise speed.
The central absorption minimum of the \CaIR\ line is typically found to be nearly at rest: for more than 4000 \CaIR\ PMJ profiles, \citet{2017A&A...602A..80D} found an average minimum position at 
0.16~\kms\ Doppler offset\footnote{This value is corrected for a conversion error in the paper.}.
Furthermore, from multiline inversions of high spatial and spectral resolution \CaK, \CaIR,\, and \ion{Fe}{i}~630~nm observations,
\citet{2019ApJ...870...88E} 
concluded that the line-of-sight plasma velocities in PMJ atmospheres do not exceed 4~\kms. 

In a study of 3D component reconnection between weak horizontal magnetic field and strong and more vertical magnetic field, \citet{2012ApJ...761...87N} 
found that 5~\kms\ jets are generated by a gas pressure gradient along the reconnecting field lines. They speculated that such a jet can be the seed of a upward-propagating slow-mode wave that becomes a shock in the upper chromosphere with a velocity amplitude of 50--100~\kms.

In this paper, we investigate the temporal evolution of PMJs with high temporal cadence \CaH\ data from the Swedish 1-meter Solar Telescope 
\citep[SST, ][]{2003SPIE.4853..341S}. 
With a cadence of 1~s, these time series are significantly faster than the 8~s cadence \Hinode\ data of the 
\citep{2007Sci...318.1594K} 
PMJ discovery paper.

\section{Observations and data reduction}
\label{sec:observations}

\begin{figure*}[!ht]
\includegraphics[width=\textwidth]{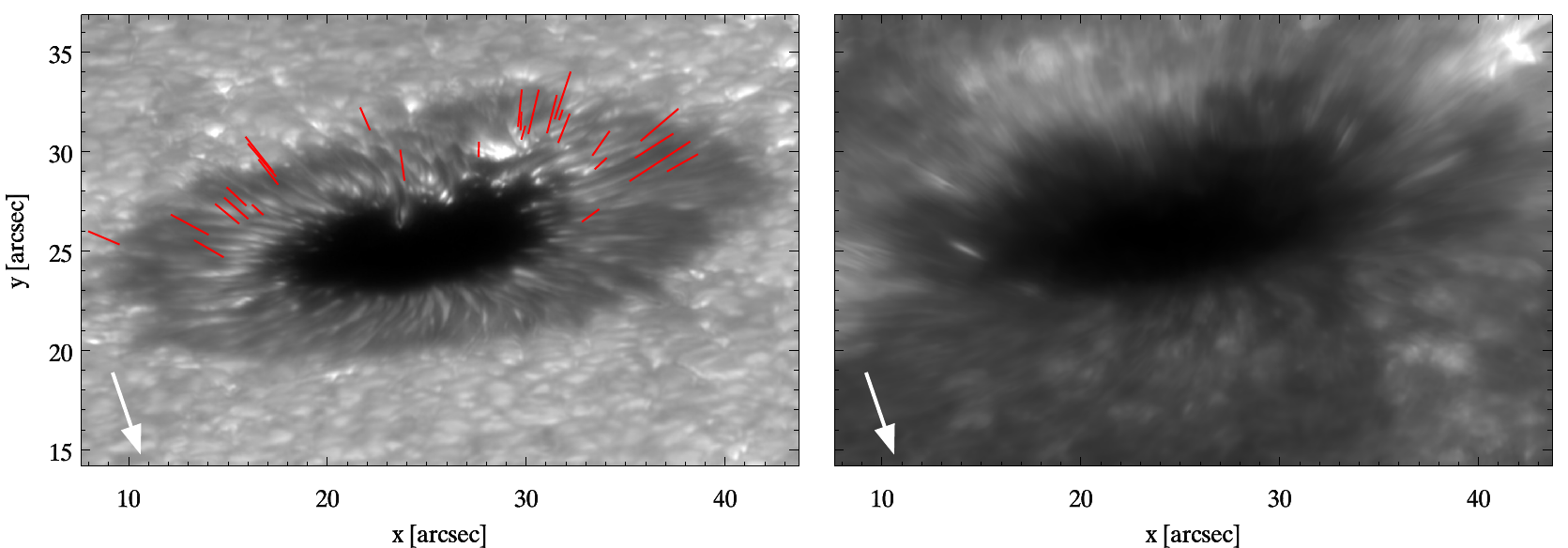}
\includegraphics[width=\textwidth]{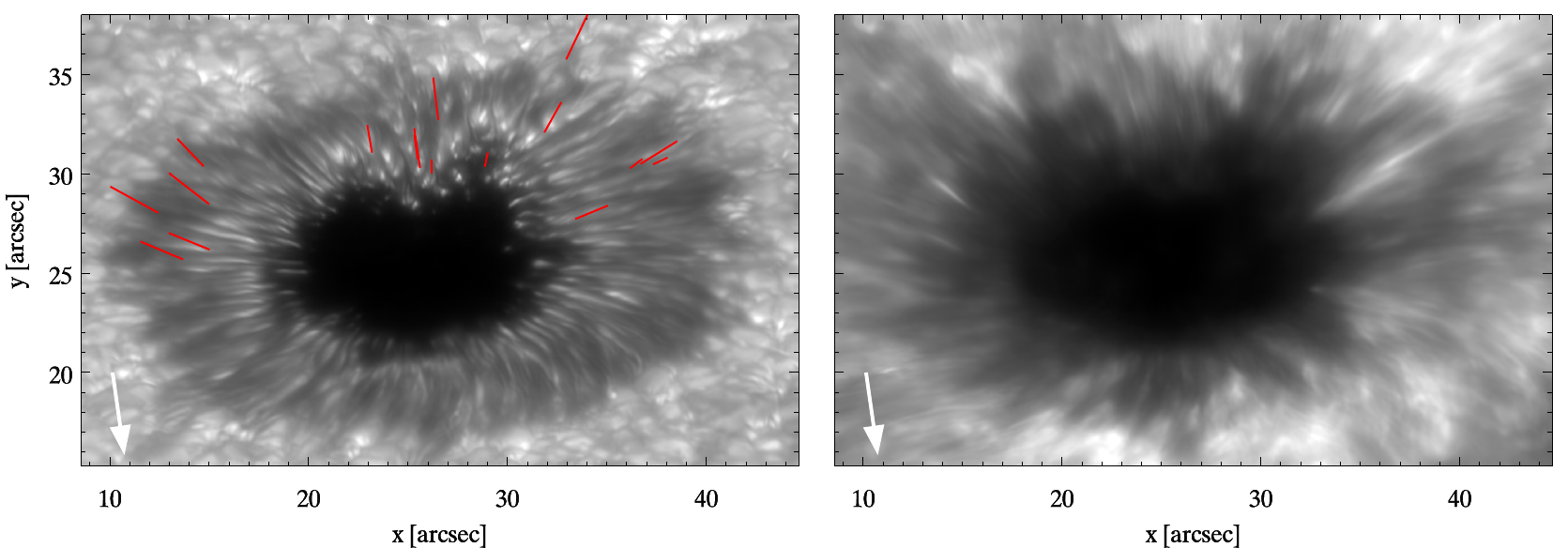}
\caption{\label{fig:overview}%
Observed sunspot in AR11084 on 27 June 2010 (top row) and 28 June 2010 (bottom row). The left panels show wideband images ($\lambda=395.4$~nm), the right panels show cotemporal \CaH\ 396.8~nm images. The white arrows in the lower left point toward disk center. The red lines in the left panels mark the extent of PMJs throughout their lifetime.
}
\end{figure*}

The observations were obtained with the SST
\citep[ ][]{2003SPIE.4853..341S} 
on the island of La Palma, Spain. 
The main sunspot in active region AR11084 was observed on 27 and 28 June 2010 at solar heliocentric coordinates $(x,y)=(-820\arcsec,-330\arcsec)$ under observing angle $\mu=\cos(\theta)=0.25$ (27 June) and at $(-720\arcsec,-343\arcsec)$, $\mu=0.53$ (28 June). 
Here we concentrate on data acquired on the blue branch of the optical path, where we used a pair of synchronized cameras, one equipped with a \CaH\ filter 
\citep[centered at the line core at $\lambda=396.85$~nm and bandpass FWHM=0.1 nm,][]{2011A&A...533A..82L}, 
and the other with a wideband filter with FWHM=1~nm, centered between the \CaH\ and \CaK\ lines at $\lambda=395.4$~nm, that imaged the solar photosphere. 
Figure~\ref{fig:overview} shows sample images of the sunspot. The \CaH\ images for both days show clear PMJ examples 
in the penumbra on the left side of the images.
Data from the red branch, spectral scans of the \CaIR\ line with the CRisp Imaging SpectroPolarimeter (CRISP) 
\citep{2008ApJ...689L..69S}, 
were analyzed by
\citet{2017A&A...602A..80D}. 
High image quality resulted from the excellent seeing conditions, the adaptive optics system
\citep{2003SPIE.4853..370S}, 
and image restoration with the multi-object multi-frame blind deconvolution 
\citep[MOMFBD, ][]{2005SoPh..228..191V} 
method. 
The blue cameras operated at a rate of 10.8~frames~s$^{-1}$ , and the data were MOMFBD-restored to a time series with a cadence of 1.02~s by including a set of 11 exposures in \CaH\ and wideband per restoration. 
For both days, the spatial resolution in the restored time series frequently approached the telescope diffraction limit of $\lambda/D$=0\farcs08 (the spatial sampling is $0\farcs034$ (or 25~km) pixel$^{-1}$).
After MOMFBD image restoration, the images were rigidly aligned and destretched to a coherent time series
\citep{1994ApJ...430..413S}. 
For 27 June 2010, the time series has a duration of 01:18:14 h, starting at 08:58:24 UT,
and for 28 June 2010, the time series has a duration of 00:41:42, starting at 09:18:29 UT.

In addition to the main 1~s cadence \CaH\ data sets, we analyzed cotemporal CRISP \CaIR\ and CHROMospheric Imaging Spectrometer (CHROMIS) \CaK\ spectral scans acquired on 3 September 2016 to address differences in detecting PMJs with the 0.3~nm \Hinode\ filter and the 0.1~nm SST filter.
The leading part of active region AR12585 at ($-568\arcsec, 420\arcsec$), $\mu=0.8$,
was observed, including a large portion of the limb-side penumbra of the main leading spot.
In this part of the penumbra, several clear PMJs were observed in both the \CaK\ and \CaIR\ lines. 
CRISP sampled the \CaIR\ line at 21 line positions ($\pm$60~\kms, with denser sampling in the core region), and CHROMIS sampled the \CaK\ line at 22 line positions ($\pm$101~\kms, with steps of 6~\kms\ out to 54~\kms). 
These CRISP and CHROMIS spectral scans were processed with the standard SST reduction pipelines
\citep{2015A&A...573A..40D, 
2018arXiv180403030L}, 
including MOMFBD image restoration.   
For more details on these data sets, we refer to 
\citet{2017ApJ...851L...6R} 
and
\citet{2019ApJ...870...88E}. 

\section{Methods}
\label{sec:methods}

In order to minimize uncertainties due to seeing variations, we concentrate our analysis on PMJs during the best seeing periods in our data sets. 
PMJs were identified visually as transient linear features that clearly display enhanced brightness relative to their surroundings. 
Efficient identification was achieved with the data exploration tool CRISPEX
\citep{2012ApJ...750...22V}. 
To analyze their temporal evolution, we extracted space-time diagrams along linear trajectories aligned with the main PMJ axis. 
We find that most PMJs display both minimal sideways motion and minimal curvature, but to account for sideways motion and misalignment, we averaged the space-time diagram over a lateral width of 3 pixels (equivalent to 75~km). 
The data for the space-time diagrams were extracted such that the end closest to the umbra was taken as the starting point, that is, all paths are oriented radially outward relative to the umbral center. 
In the discussion of the results, we refer to the direction away from the umbra as ``up'' and toward the umbra as ``down''.
The reference time $t=0$~s is defined for each PMJ at the time of maximum intensity in the space-time diagram.
The measurements of the PMJ length and width were taken at this time, and the maximum intensity served as the reference for the brightness increase (in~percent).

As a measure of the seeing variation during the PMJ evolution, the figures presenting the space-time diagrams include a panel with the contrast in the cotemporal (photospheric) wideband images. 
The contrast is measured as the standard deviation divided by the mean of a relatively quiet area away from the sunspot.
While contrast in the granulation pattern certainly does not capture all effects of seeing on image quality, it serves as an indicator of the general quality of the seeing. 

The PMJs in our sample display a wide variety in dynamical evolution, and the presentation of our results requires considerable figure space to highlight different aspects of the evolution. 
For the sake of clarity, we have tried to minimize the number of PMJ examples in the main body of the article. To ensure completeness of the presentation, we include a number of additional examples of PMJ evolution in figures in the Appendix, to which we refer in the Results and Discussion sections. 

\section{Results}
\label{sec:results}

We selected 45 PMJs for further analysis (28 in the 27 June 2010 data set and 17 in the 28 June 2010 data set): for these events the seeing was stable enough throughout their evolution for a proper analysis.
The PMJs in our data sets display a wide variation in dynamical evolution.
A generally common property is the rapid increase in intensity of a fibril that was already present before the onset of the PMJ. 
This brightening of the fibril occurs in a matter of seconds and is rather uniform over a significant length.
After brightening up, we observe a variety of scenarios: the top of the PMJ may rise, it may retract, or it may not move in a significant manner at all. 
The motion of the PMJ top is often the continuation of the motion of the fibril top before brightening up, and the fibril may continue with this motion after the brightening has ended and the PMJ is considered to have faded away. 
If there is PMJ top motion, the apparent speed is typically on the order of 10--15~\kms.
In some cases, the bottom of the PMJ rises toward the end of its lifetime. %
For PMJs with a rising motion of the top, this results in an apparent proper motion of the PMJ. 
For PMJs without a rising top, the rising of the PMJ bottom results in an apparent shrinking. 
We also observe an apparent splitting of the PMJ for a number of cases. 

In the following paragraphs, we discuss a number of PMJs in detail. We also illustrate the various PMJ evolution scenarios. 

\begin{figure*}[!th]
\includegraphics[width=\textwidth]{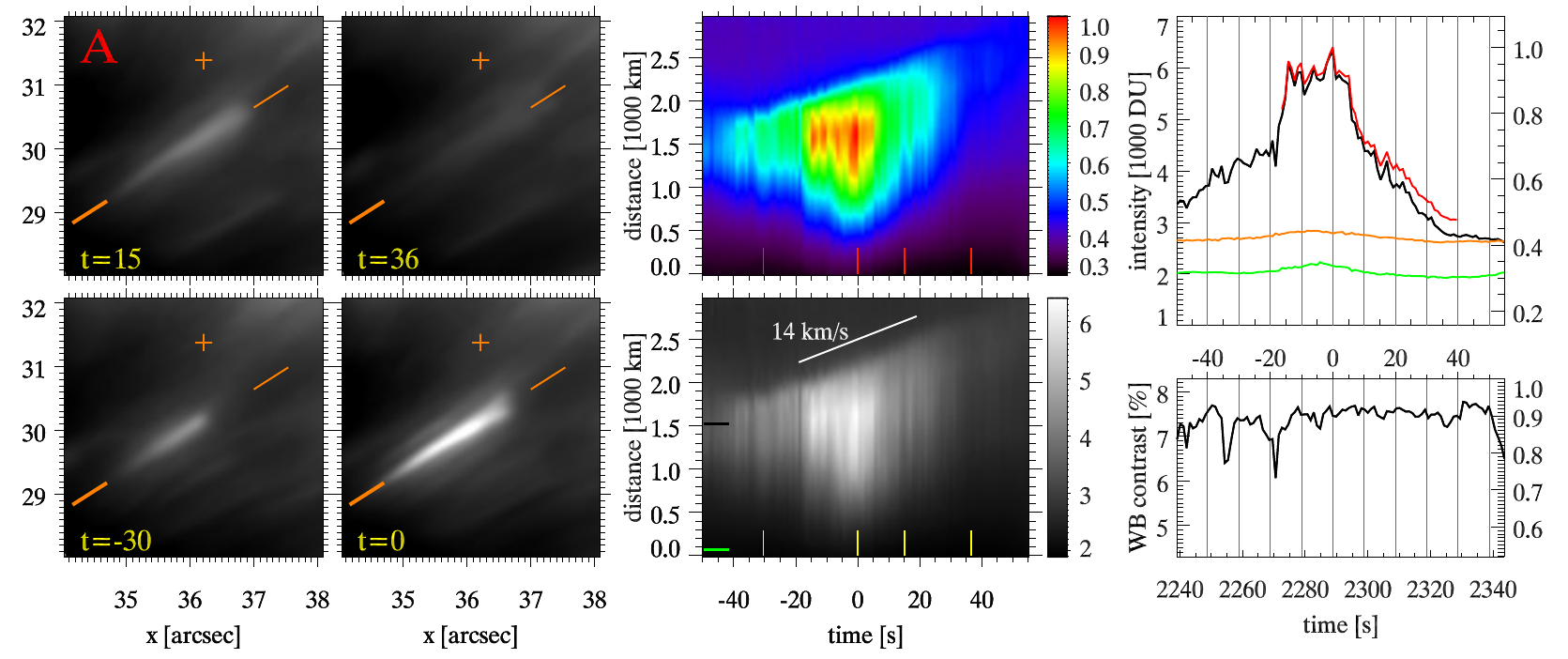} \\
\includegraphics[width=\textwidth]{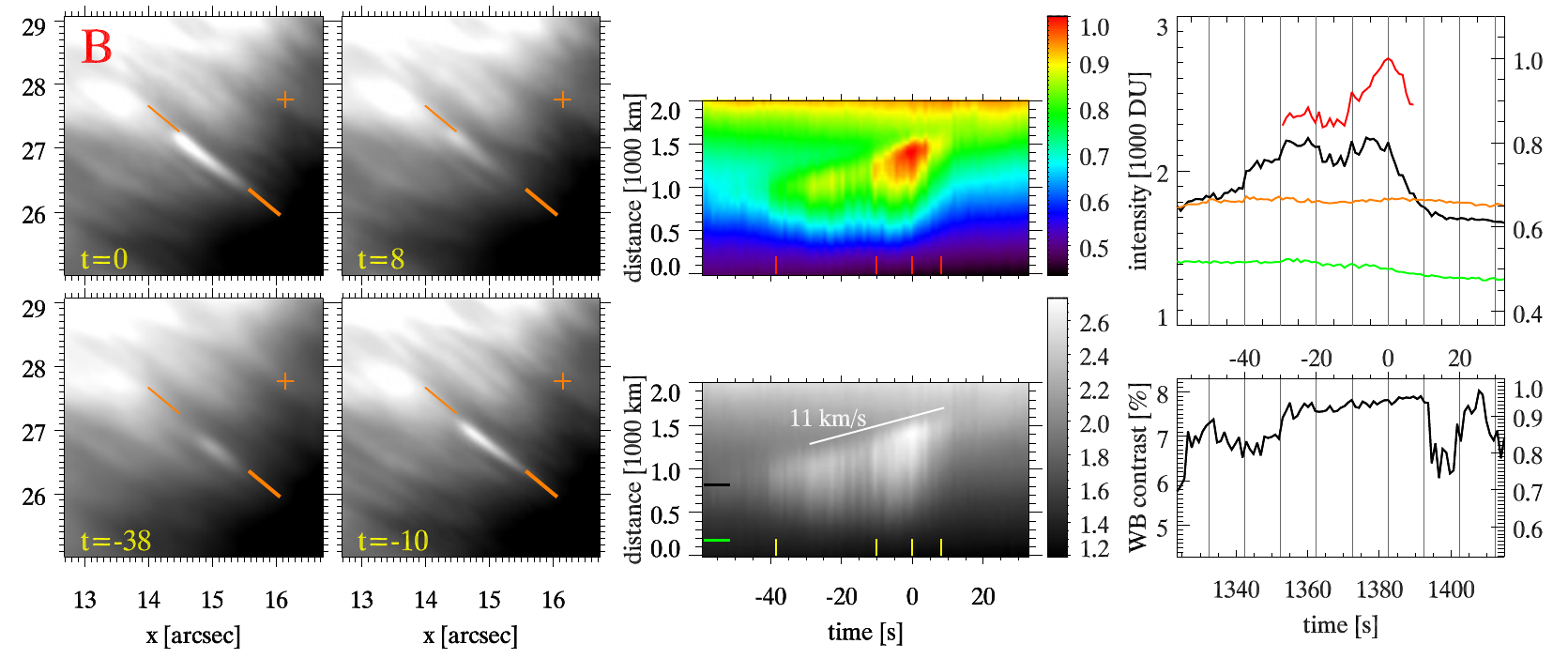}
\caption{\label{fig:detailsAB}%
Details of the evolution of two PMJs labeled A and B. The left four images show selected moments of the area centered on the PMJ. The time (in seconds) is marked in the lower left corner and is relative to the time when the maximum PMJ signal is measured in the space-time diagram. The four images are all scaled to the PMJ maximum at $t=0$ according to the gray-scale color bar. The start and end points of the trajectory of the space-time diagram are indicated with orange markers, the thicker marker indicates the start and is closest to the umbra. Two versions of the space-time diagram are shown with two different color tables: in gray scale and rainbow-color scale. A white line indicates the slope of a specified propagation speed. Light curves are shown in the upper right panel; the black and green lines show the intensity at fixed locations in the space-time diagram. These locations are marked in the gray-scaled space-time diagrams. The red line shows the highest PMJ signal during the brightest phase of the PMJ. The orange line shows as reference the intensity in the vicinity of the PMJ but outside the space-time diagram. Its location is marked with the orange cross in the images on the left. The units are arbitrary detector units. The right axis scaling is normalized to maximum PMJ intensity (at time t=0 s). In the lower right corner the intensity contrast in the wideband channel is shown as an indication of the seeing quality. The right axis scaling is normalized to maximum contrast in the whole time series. Animations of this figure are available at \url{http://folk.uio.no/rouppe/pmj_highcadence/}. 
}
\end{figure*}

Figure~\ref{fig:detailsAB} shows details of the evolution of two PMJs with a rising top.
For PMJ~A, at time $t=-17$~s, the brightness of a rather faint fibril starts to rise quickly and increases with more than 20\% of the PMJ peak intensity in only 3~s. 
In only 1~s, from $t=-15$ to $t=-14$~s, the PMJ has brightened over the 90\% threshold over a length of 370~km, which is visible in the top colored space-time diagram as the sharp transition from green to red.
PMJ~A reaches maximum brightness at $t=0,$ at which time the PMJ is above 90\% brightness over a length of 670~km and the FWHM of the PMJ measures 1270~km. 
From $t=5$~s, the brightness starts to drop quickly, and in 2~s, the brightness has dropped to below 80\%.
A fibril continues to be visible for at least 20 more seconds, after which the intensity drops to below 60\%. 
During this whole period, the top of the PMJ is moving upward (away from the umbra) with a velocity of about 14~\kms, as indicated by the inclined white line in the bottom space-time diagram. 
After $t=0$, the bottom of the PMJ starts to move upward with roughly twice that speed so that the PMJ appears to be shrinking.
%

Penumbral microjet~B displays a similar evolution, with a faint fibril brightening up for a limited duration of time, and at the same time, the PMJ top moves at $\sim11$~\kms. 
During the decay of the PMJ, the bottom moves upward with higher speed so that the PMJ appears to be shrinking. 
Compared to PMJ~A, the rise in intensity for PMJ~B is more gradual with an increase from $\sim65$\% to $\sim80$\% over about 30~s, and another 20~s before the intensity rises above 90\%.
The brightest phase, like for PMJ~A, starts abruptly, when at $t=-10$~s, PMJ~B brightens above the 90\% threshold over a length of 200~km in only 1~s. 
At the time of maximum brightness, at $t=0$, the length of the PMJ that is over 90\% intensity is about 500~km.
After $t=5$, the intensity drops quickly below the 90\% level and continues to decline while the top of the PMJ continues to move up at $\sim11$~\kms\ and the PMJ as a whole appears to shrink.
The visibility of the full decay phase is unfortunately difficult to discern as the seeing deteriorates for a short period, as testified by the wideband contrast values. 

Penumbral microjets K and L in Fig.~\ref{fig:app_detailsKLM} are two more examples of PMJs that have an upward-moving top, with velocities 5 and 11~\kms\ , respectively. 
In total, we find that 20 PMJs in our sample (44\%) have an upward-moving top during the bright phase of the PMJ. 

\begin{figure*}[!th]
\includegraphics[width=\textwidth]{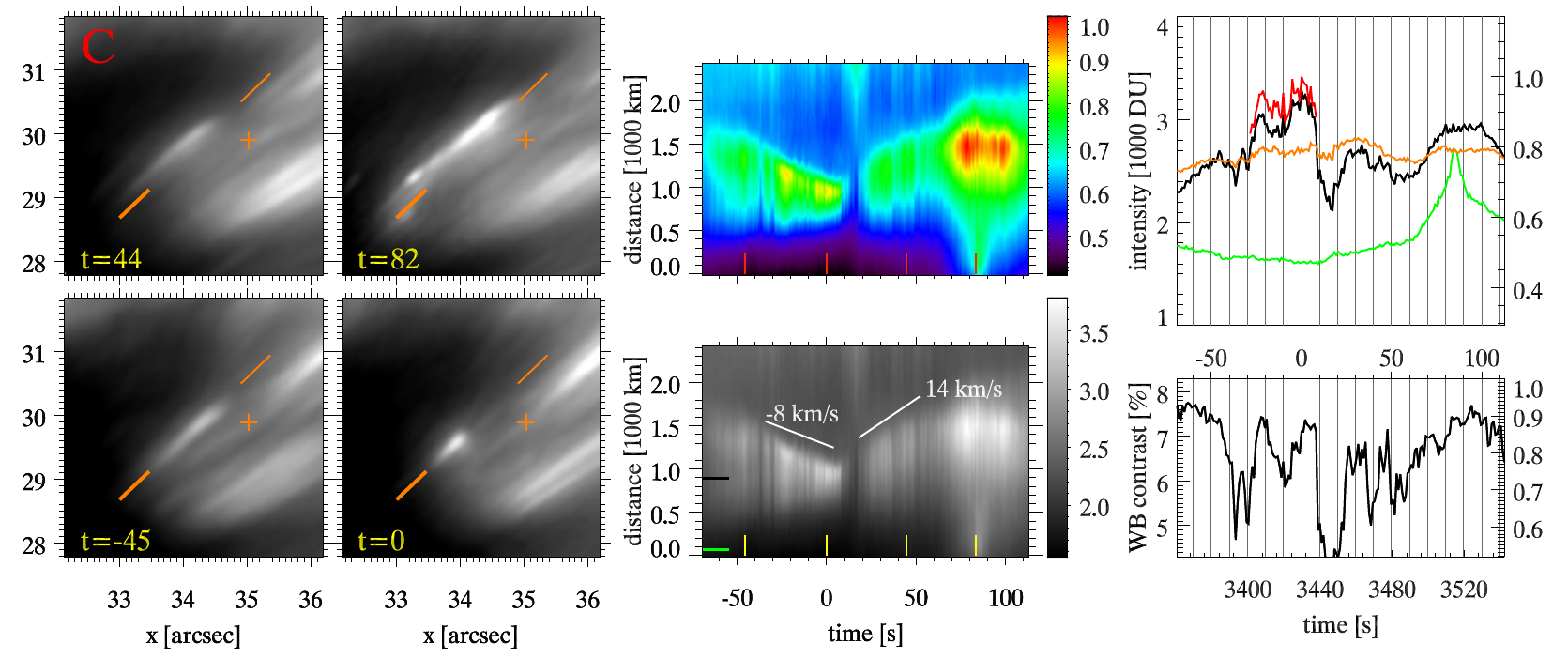} \\
\includegraphics[width=\textwidth]{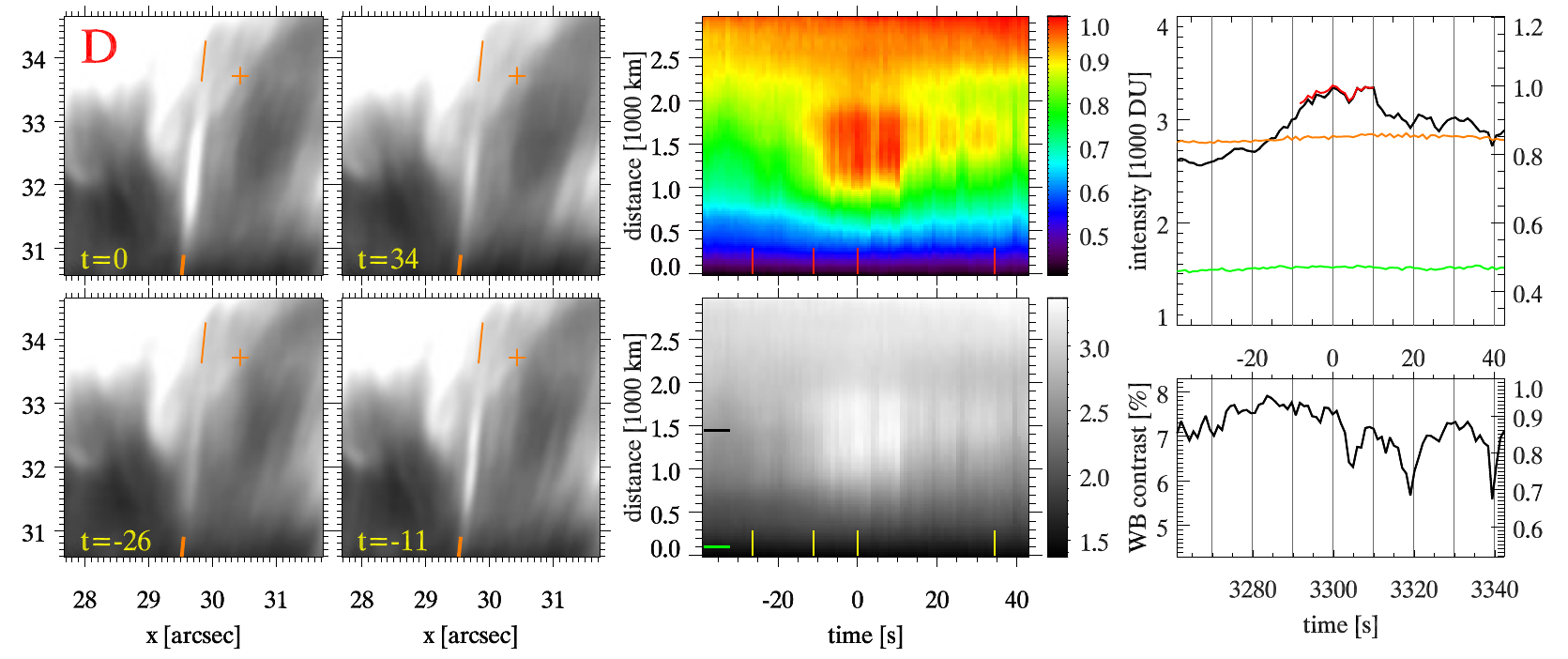}
\caption{\label{fig:detailsCD}%
Details of the evolution of PMJ~C with a retracting top (top) and PMJ~D that does not display motion of the top (bottom). The format of this figure is the same as that of Fig.~\ref{fig:detailsAB}. Animations of this figure are available at \url{http://folk.uio.no/rouppe/pmj_highcadence/}.
}
\end{figure*}

The top of PMJ~C in Fig.~\ref{fig:detailsCD} moves downward, and the overall impression is that of a PMJ that shrinks over a time period of about 20~s. 
The apparent velocity of the moving top is about 8~\kms. 
At the end of the PMJ, just when the PMJ seems to vanish, the seeing becomes very poor and the actual moment of disappearance is completely blurred out. 
When the seeing improves again after about 15~s, a new PMJ seems to have appeared at a position slightly to the side of the position where the short PMJ C vanished earlier.
The top of this new PMJ rises with an apparent speed of about 14~\kms. 

Penumbral microjet~D in Fig.~\ref{fig:detailsCD} appears not to be moving at all. 
A fibril with an intensity of about 80\% of PMJ maximum suddenly starts to brighten up over a significant length: it rises above 90\% over a length of 650~km at $t=-11$ in only 4~s (with a jump from 50~km to 650~km in only 1~s).
At maximum brightness ($t=0$), the length above 90\% is 1145~km. 
The bright phase above 90\% lasts for about 46~s, where the seeing conditions are worse in the last about 30~s. 
During this whole period, neither the top nor the bottom of the PMJ seem to be moving, so that the PMJ appears to be stationary. 
We find that the tops of 3 PMJs move downward (7\%) and that the tops of 22 PMJs have no clear motion (49\%).
An absence of clear motion of the top implies an upper limit of about 2~\kms.

\begin{figure*}[!th]
\includegraphics[width=\textwidth]{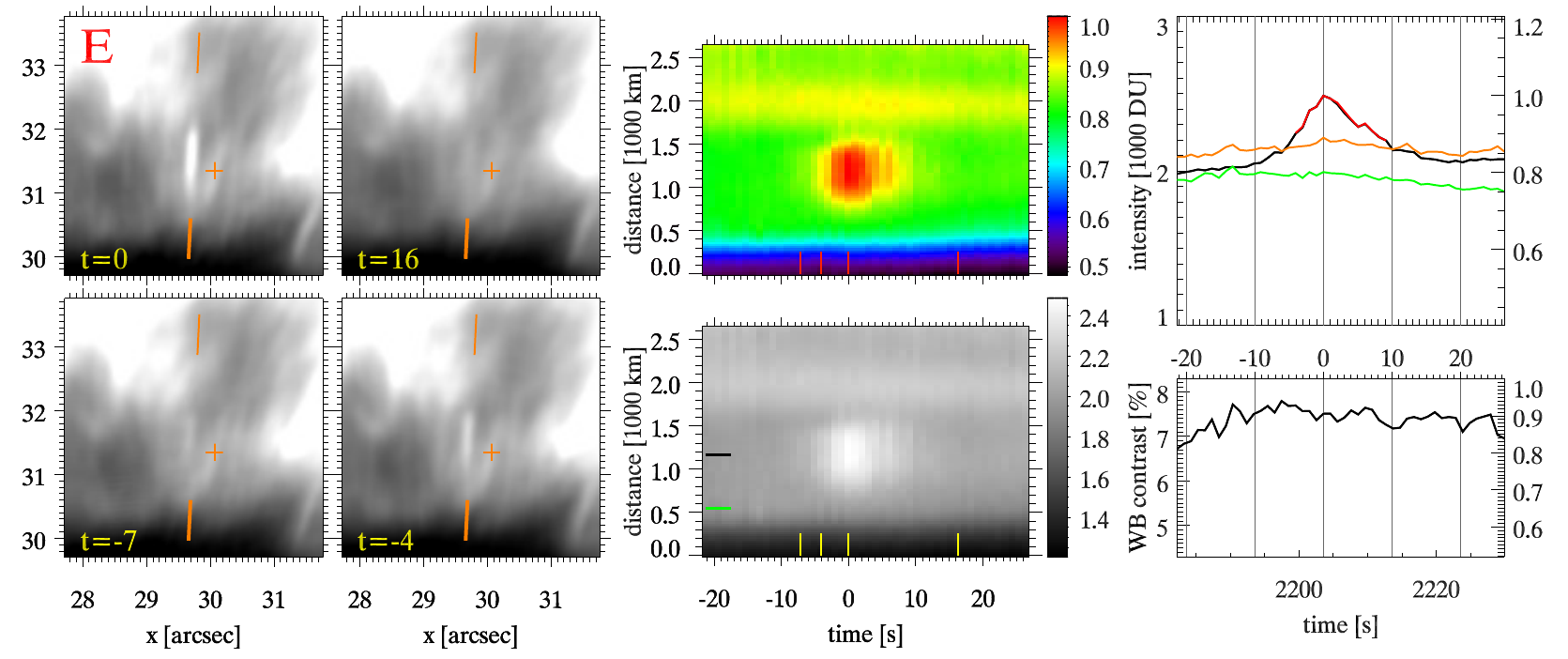} \\
\includegraphics[width=\textwidth]{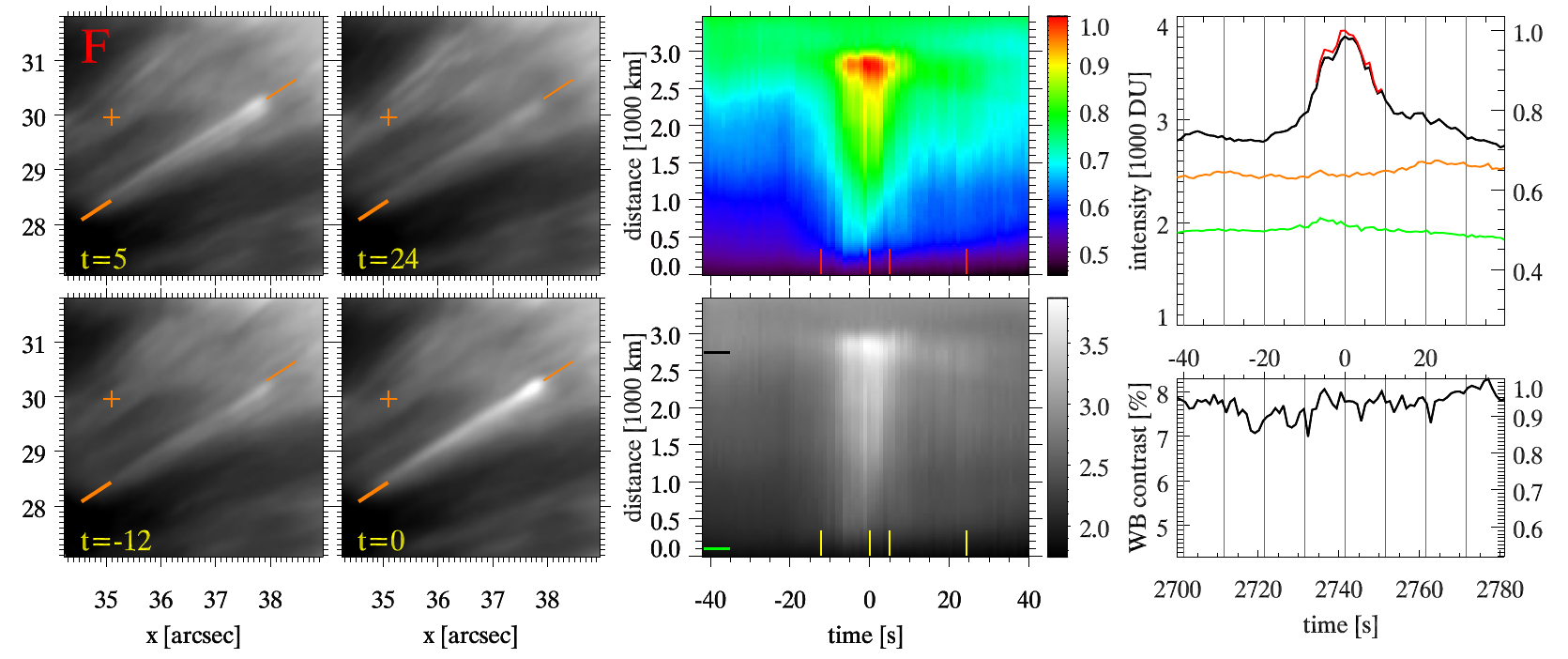}
\caption{\label{fig:detailsEF}%
Details of the evolution of two PMJs with short lifetimes. PMJ~E is short (top) and PMJ~F is long (bottom). The format of this figure is the same as that of Fig.~\ref{fig:detailsAB}. Animations of this figure are available \url{http://folk.uio.no/rouppe/pmj_highcadence/}.}
\end{figure*}

The tops of PMJs E and F in Fig.~\ref{fig:detailsEF} also display no apparent motion. 
Both are short-lived events (11 and 12~s above 90\%), and while PMJ~E appears like a mere stationary brightening of the fibril, PMJ~F appears to be growing quickly toward the bottom and then to shrink as quickly from the bottom up. 
The relative intensity of PMJ E increases by about 20\%, and that of PMJ~F by about 30\%.
%

\begin{figure*}[!th]
\includegraphics[width=\textwidth]{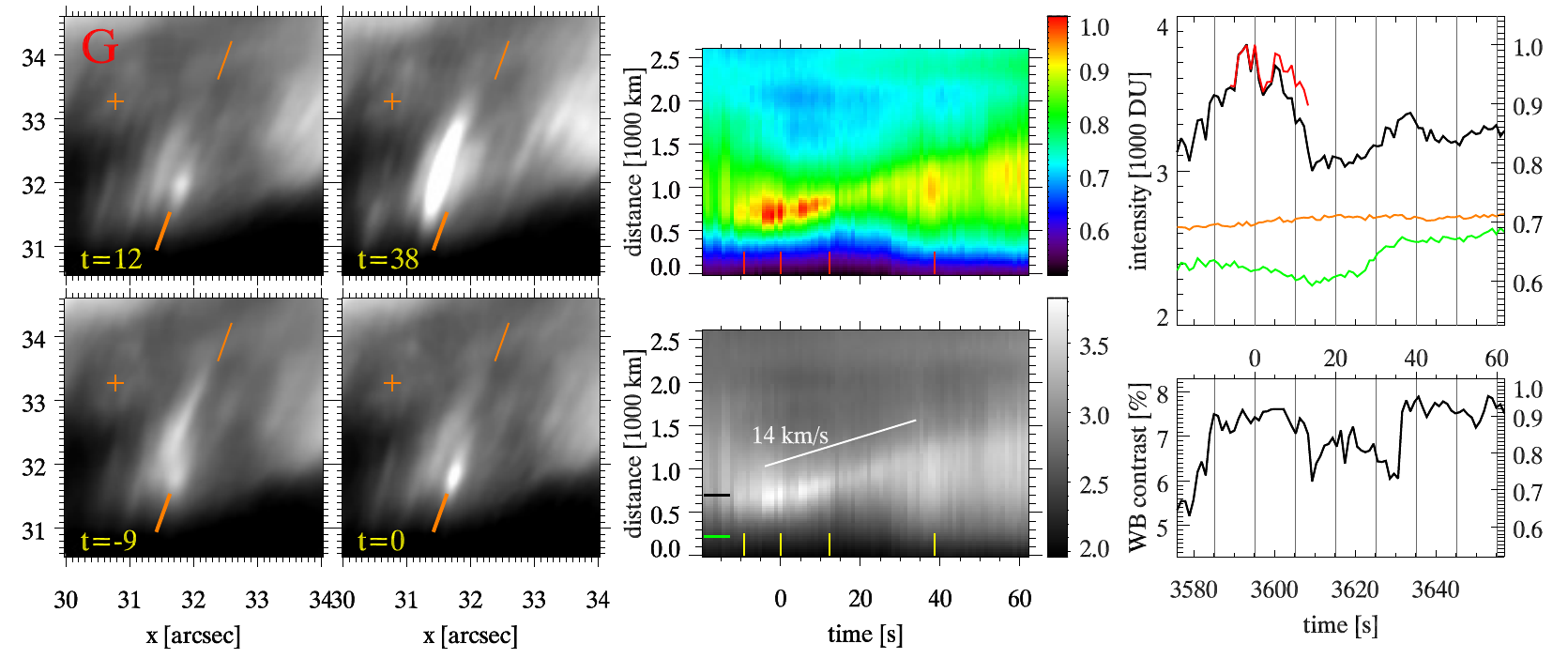} \\
\includegraphics[width=\textwidth]{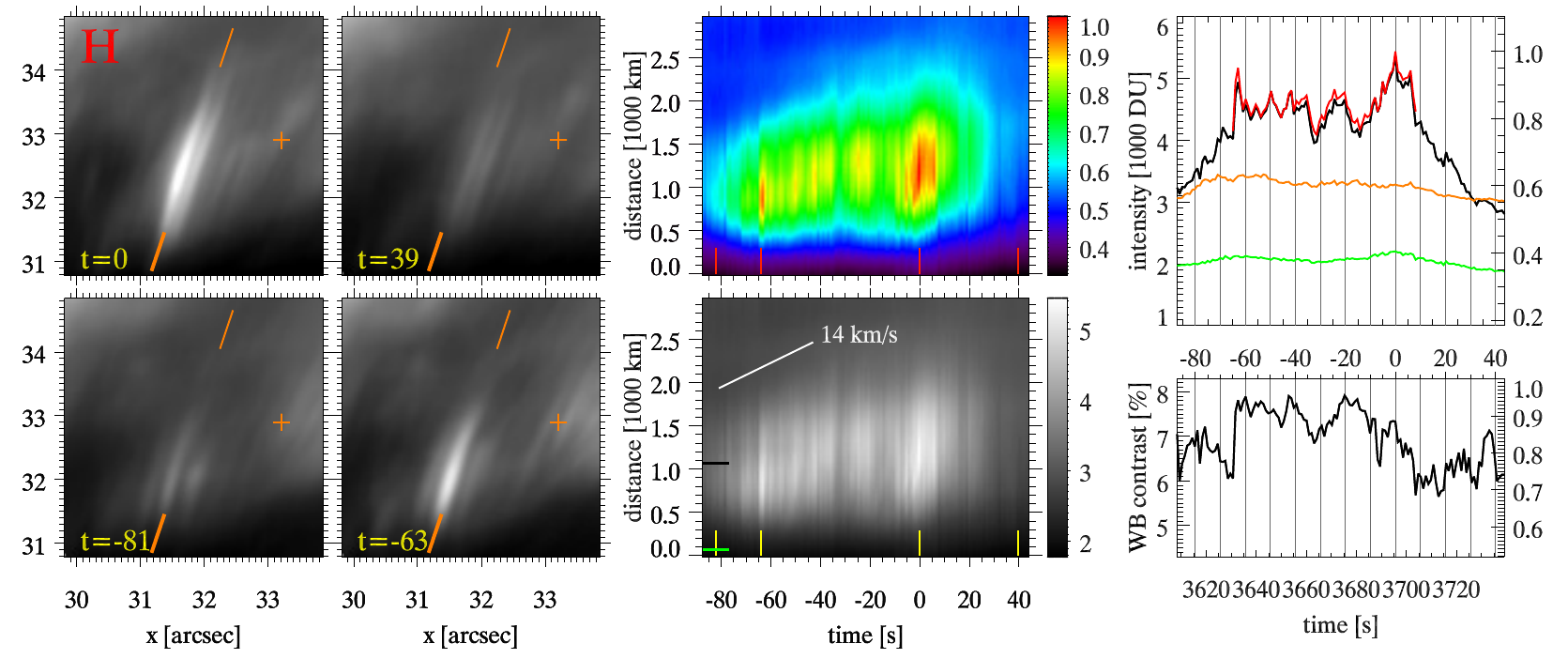}
\caption{\label{fig:detailsGH}%
Details of the evolution of two PMJs that occurred in close proximity after one another. Maximum brightness of PMJ~G ($t=0$) occurred 99~s before the maximum brightness of PMJ~H. In the image at $t=38$ of PMJ~G, PMJ~H can be seen to the left. This is 61~s before the maximum brightness of PMJ H (time $t=-61$). PMJ~H is a brighter and longer-lived PMJ than PMJ~G (in addition to being longer). Both PMJs appear to be splitting during the later phases of their evolution. The format of this figure is the same as that of Fig.~\ref{fig:detailsAB}.  Animations of this figure are available \url{http://folk.uio.no/rouppe/pmj_highcadence/}.}
\end{figure*}

Penumbral microjet~G in Fig.~\ref{fig:detailsGH} is an example of a PMJ with both a moving top and moving bottom end, which gives the impression of a short object moving with a velocity of 14~\kms. 
During the brightening phase, at $t=-6$, the intensity rises above 90\% over a length of 150~km in 1~s, and later, at maximum brightness at $t=0$, the length over 90\% increases from 224~km to 374~km.
PMJ~M in Fig.~\ref{fig:app_detailsKLM} is another example of a short moving object (12~\kms). In addition, PMJ~M is short-lived, with a lifetime of 11~s above 90\% of peak intensity. 

When we consider the temporal evolution of PMJ~G in detail (see the animation accompanying Fig.~\ref{fig:detailsGH}), it appears that the PMJ splits during the later phases. From about $t=38$, we can discern two parallel linear features at the location where before the PMJ was a single linear feature. 
This splitting of the PMJ can be observed in more PMJs, and a particularly clear example is PMJ~H in Fig.~\ref{fig:detailsGH}.
This PMJ splits in the later phase of the PMJ, at time $t=-9,$ which is 54~s after PMJ~H first reached intensity $>90$\%.
The double structure can be discerned long after the PMJ has decreased in intensity: at $t=39$, with the intensity below 60\%, two parallel linear features are still visible. 
Other examples of PMJ splitting are shown for PMJ~N and O in Fig.~\ref{fig:app_detailsNOP}. 
Fourteen PMJs in our sample (31\%) split into two separated fibrils at some phase in their evolution.

\begin{figure*}
\includegraphics[width=\textwidth]{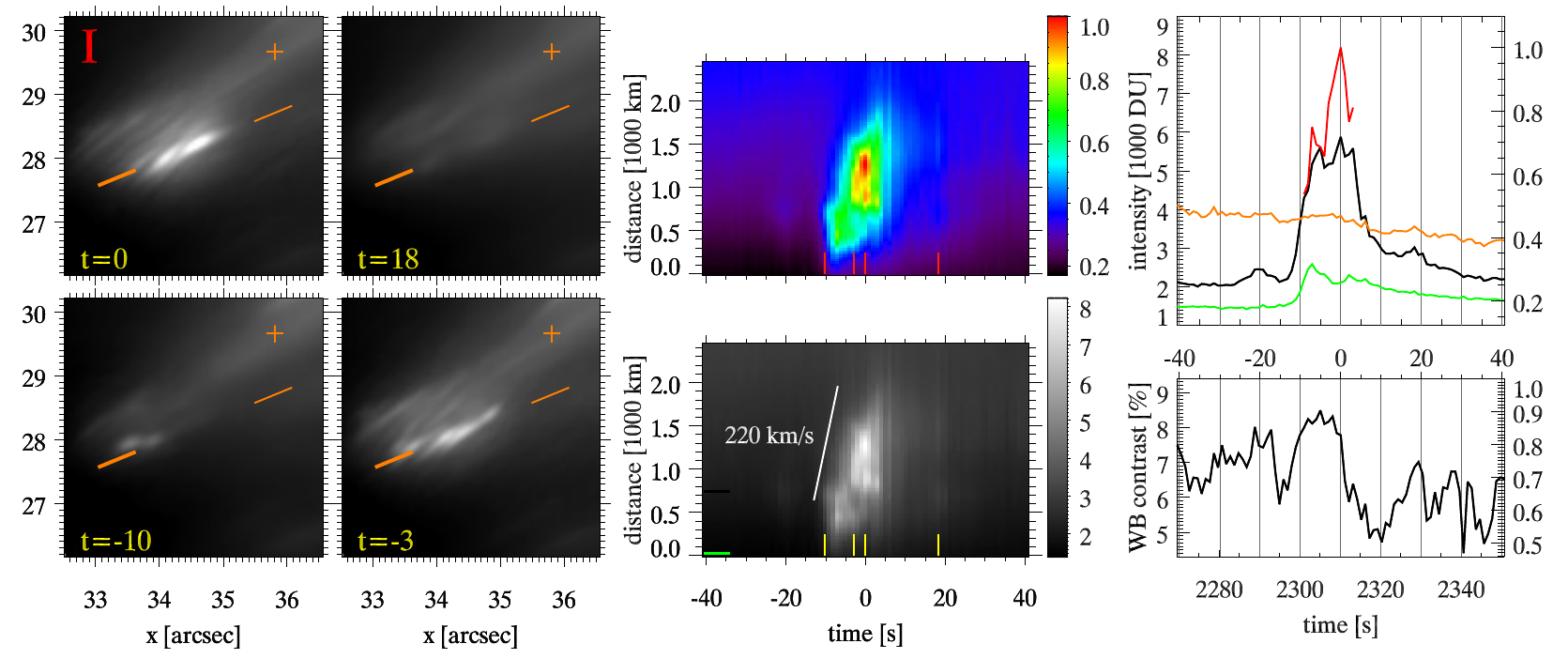} \\
\includegraphics[width=\textwidth]{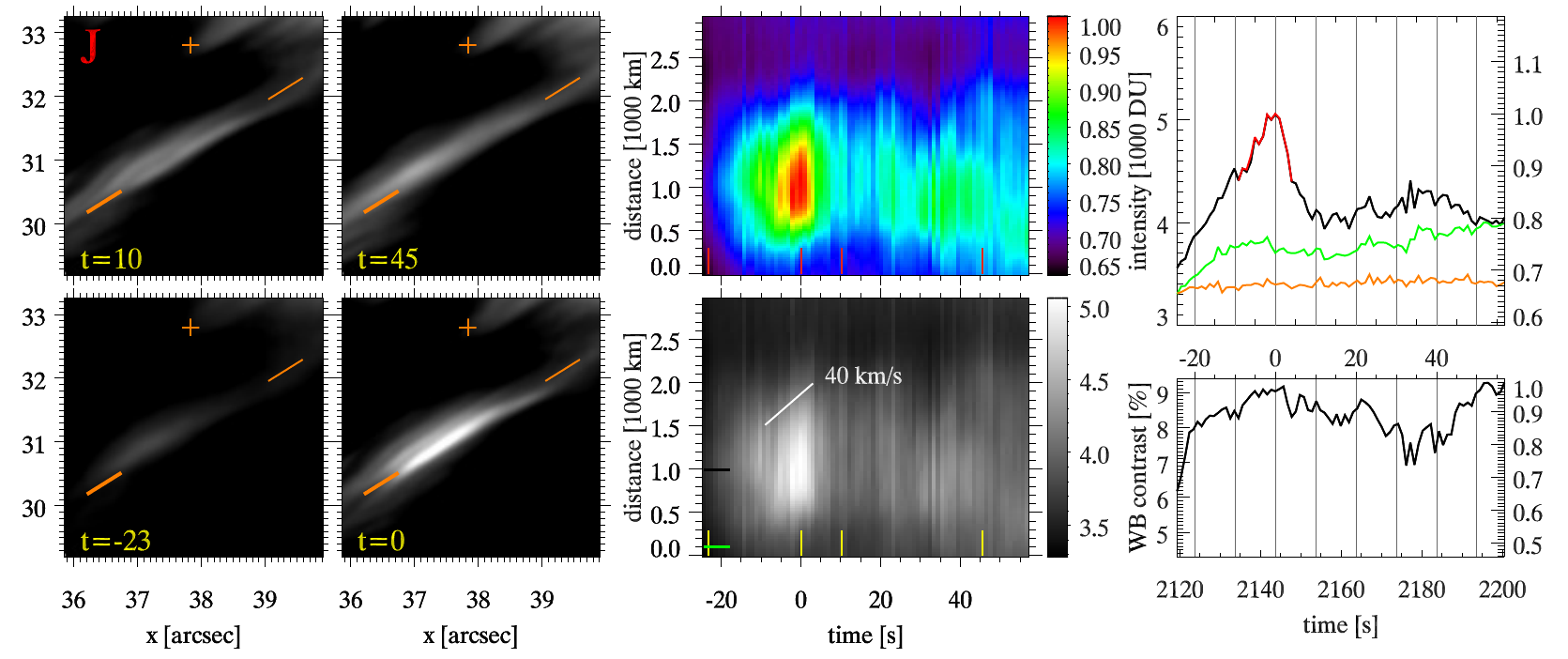}
\caption{\label{fig:detailsIJ}%
Details of the evolution of a complex event (PMJ~I, top) and an event that has a $\sim40$~\kms\ apparent rise of the top (PMJ~J, bottom). The format of this figure is the same as that of Fig.~\ref{fig:detailsAB}. Animations of this figure are available \url{http://folk.uio.no/rouppe/pmj_highcadence/}.}
\end{figure*}

Penumbral microjet~I in Fig.~\ref{fig:detailsIJ} displays an extremely rapid morphological evolution that we only observe for a few cases in our sample. 
It displays an extreme rise in intensity that is about 75\% of the peak PMJ intensity and is highest in absolute detector units compared to the other PMJs. 
Furthermore, the morphology during the bright PMJ phase varies rapidly from time step to time step, and it seems that the bright fibril breaks up into smaller clumps. 
These clumps can be seen in the images for $t=-10, -3$, and $0$ in Fig.~\ref{fig:detailsIJ}.
PMJ~I shows less of the linearly coherent evolution that we see for most other PMJs, and the question is whether this PMJ should even be regarded as a single ``standard'' PMJ event. 
The space-time diagram for this event has a pronounced slope that suggests a propagation speed of about 220~\kms\ , but the accompanying movie shows that it is questionable whether we observe an actual proper motion of a coherent structure. 
The movie also clearly shows that neighboring fibrils brighten up in phase with the bright clumps of the event. 

A similar occurrence of a complex pattern of small blobs along the PMJ fibril can be seen in the late evolution of the PMJ that appeared after the disappearance of PMJ~C in Fig.~\ref{fig:detailsCD}: the accompanying movie shows small, downward-moving blobs from $t=79$.
We also note that both PMJ~C and I are located in a region of the penumbra that was marked as a ``hot spot'' of PMJs in 
\cite{2017A&A...602A..80D}, 
a region where many and sometimes overlapping PMJs occurred. PMJs in this hot spot often had particularly pronounced \CaIR\ spectral profiles. 

Penumbral microjet~J in Fig.~\ref{fig:detailsIJ} is also located in the hot-spot area and displays an apparent motion of the top of the PMJ that is significantly faster than for the other PMJs: $\sim40$\kms. 
The apparent evolution of PMJ~P in Fig.~\ref{fig:app_detailsNOP} is affected to some extent by seeing variation, but PMJ~P seems to display faster upward motion than most PMJs. At some phase in the evolution, around $t=38$, there is an apparent rise speed of roughly 140~\kms. 

\begin{figure*}
\includegraphics[width=\columnwidth]{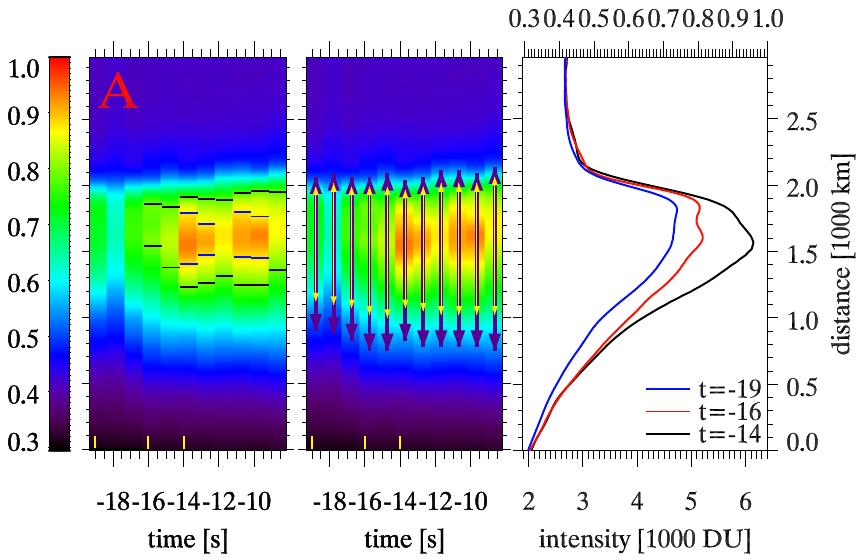}
\includegraphics[width=\columnwidth]{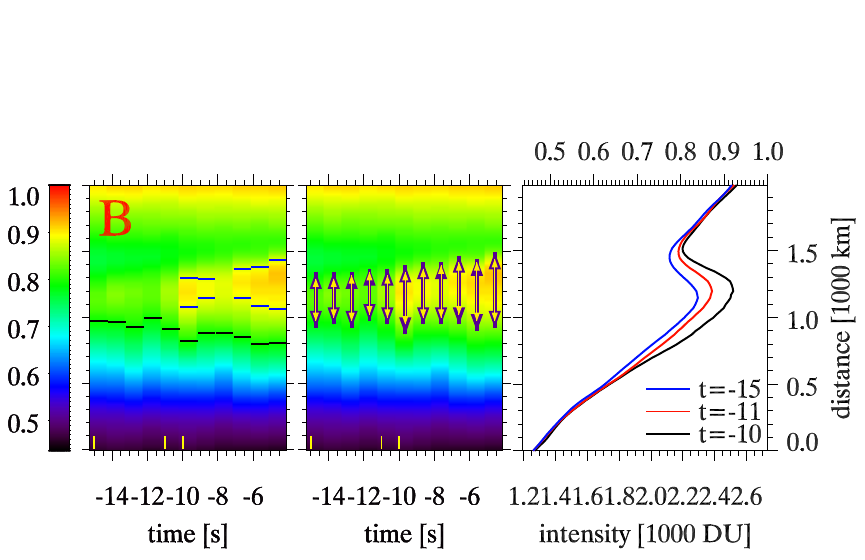} \\
\includegraphics[width=\columnwidth]{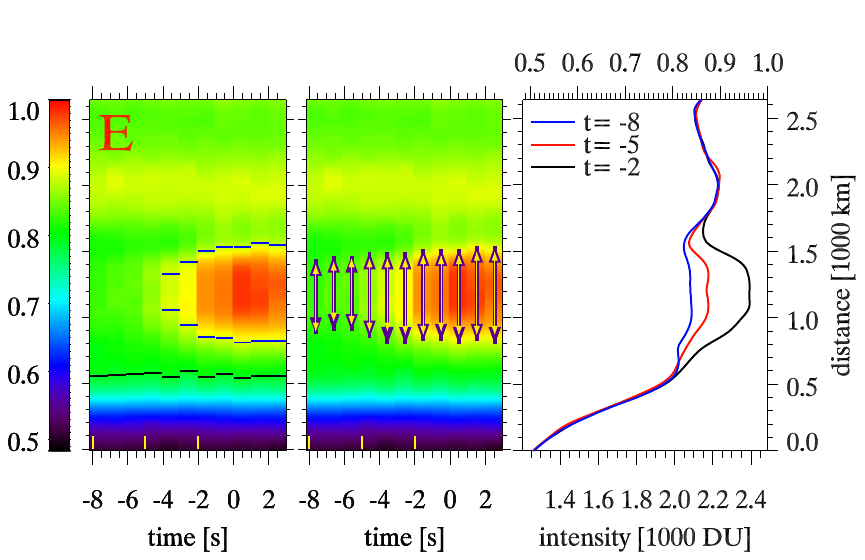}
\includegraphics[width=\columnwidth]{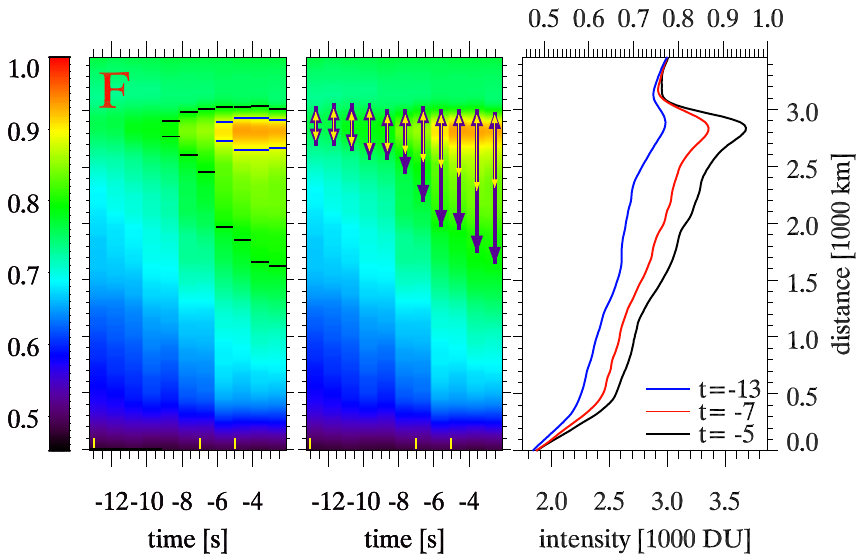} \\
\caption{\label{fig:onset}%
Detailed look at the onset of PMJs. 
Two identical space-time diagrams are shown, with the color-scaling (shown at left) relative to the maximum intensity of the PMJ at $t=0$. In the left diagram, horizontal blue and black lines serve as contours at the 90\% and 80\% intensity levels, respectively, relative to the maximum of the PMJ. In the right space-time diagram, the vertical lines with arrowheads show the full width at 25\% (FWQM, thick purple) and 50\% (FWHM, thin yellow) of the maximum intensity at that time step. The right panel shows the intensity profiles for three selected time steps. 
Time is relative to the maximum intensity of the PMJ (t=0~s).
The PMJs are identified by the letters in the upper left corner, which refer to PMJ~A and B in Fig.~\ref{fig:detailsAB} and PMJ~E and F in Fig.~\ref{fig:detailsEF}.
}
\end{figure*}

A more detailed look at the onset of PMJs is provided by Fig.~\ref{fig:onset} , which shows details over 11~s around the time when PMJs A, B, E, and F suddenly brighten. 
Two identical space-time diagrams are shown that zoom-in on the longer-duration space-time diagrams of Figs.~\ref{fig:detailsAB} and \ref{fig:detailsEF}.
These space-time diagrams have two different types of markers that serve as length indicators: the left diagram has horizontal contour markers of the 80\% and 90\% levels of the peak PMJ intensity at $t=0$, and the right diagram has vertical lines that show the extent of the full-width at half-maximum (FWHM) and full-width at quarter-maximum (FWQM) with the maximum measured for each time step. 
From the FWHM and FWQM markers it is clear that a fibril was present at the space-time path before the sudden increase in intensity. 
This is also illustrated in the profile panel at the right where the blue profile is from a few seconds before the onset. 
The blue profiles have similar shapes as the black onset profiles, it is only at lower intensity. 
The PMJ rises in intensity basically over the full length of the fibril, that is to say, there is no clear  high-velocity upward motion from the bottom up, a motion that can be expected from a jet.
PMJ~A rises over the 80\% level over a length of 350~km in 2~s (from $t=-16$ to $t=-14$) and rises over the 90\% level over 374~km in 1~s.
PMJ~B rises over 90\% over 200~km in 1~s, and for PMJ~E, the length over 90\% increases to 622~km in 3s.
PMJ~F rises over 90\% over 150~km in 1~s, and this length increases to 250~km in the next time step. During the same 2~s, the length over 80\% increases with about 620~km to 1145~km. 
Here, the increase in length is mostly from the top down, and the bottom 80\% marker in the left space-time diagram moves with a velocity of about 220~\kms, while the bottom of the FWQM markers increases with a velocity of about 150~\kms. 
The bottom of the 90\% level of PMJ~D in Fig.~\ref{fig:app_onset}  grows at an apparent speed of 112~\kms.

Penumbral microjet~P in Fig.~\ref{fig:app_onset} zooms in on the phase that has an apparent rise of about 140~\kms\ (as indicated in Fig.~\ref{fig:app_detailsNOP}). Here, the top of the 90\% level rises with a mere 25~\kms. 
The apparent rise of the top of PMJ~J in Fig.~\ref{fig:detailsIJ}  is 40~\kms. Zooming-in on the onset in Fig.~\ref{fig:app_onset} shows that the top of the 90\% level moves with 42~\kms. 
Zooming-in on the complex PMJ~I in Fig.~\ref{fig:app_onset} shows the fine structure in this event, for which it is questionable whether we can measure a rise velocity similar to that of other PMJ events.  

\begin{figure*}
\includegraphics[width=\textwidth]{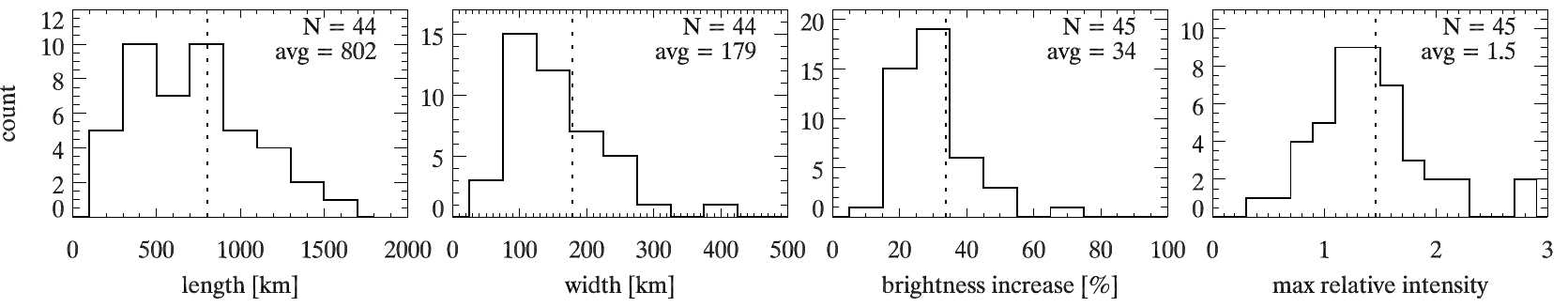}
\caption{\label{fig:stats} %
Histograms for the lengths, widths, brightness increase, and maximum relative intensity for the PMJs we analyzed. Length and width are measured as the FWHM at the time of maximum PMJ brightness. For PMJ~I, no meaningful length and width could be measured. The brightness increase is measured as the intensity increase at a fixed position in the space-time diagram from the minimum to the maximum PMJ intensity. The fixed position is the black line in Figs.~\ref{fig:detailsAB}--\ref{fig:detailsIJ}. The maximum PMJ intensity is relative to the average \CaH\ intensity in a relatively quiet area outside the sunspot.
}
\end{figure*}

Figure~\ref{fig:stats} shows histograms of the lengths, widths, brightness increase, and maximum relative intensity for the 45 PMJs in our sample. 
Length and width are measured as the FWHM at the time of maximum PMJ intensity ($t=0$). 
The average FWHM length is 802~km (median 818~km) and the average FWHM width is 179~km (median 170~km).
The sharp drop for widths below 100~km is likely due to the spatial resolution of the telescope (diffraction limit equivalent to 60~km).
The brightness increase is measured at a fixed location in the space-time diagram (marked with the black line in Figs.~\ref{fig:detailsAB}--\ref{fig:detailsIJ}) and is the difference between the minimum at this location and the maximum intensity of the PMJ during its lifetime (i.e., at the time defined as $t=0$).
The average brightness increase is 34\% (median 32\%). 
The maximum relative intensity is measured as the maximum PMJ intensity at $t=0$ relative to the average \CaH\ intensity in a relatively quiet area outside the sunspot (57\arcsec $\times$ 16\arcsec\ for 27 June 2010 and 57\arcsec $\times$ 13\arcsec\ for 28 June 2010).
The average maximum relative intensity is 1.5 (median 1.4). 
The maximum intensity of most PMJs is higher than the average intensity in the quiet regions outside the sunspot.

\section{Discussion}
\label{sec:discussion}

The most notable result from our analysis of 1~s cadence \CaH\ sunspot time series is that PMJs appear as rapid brightenings of existing, faint fibrils. 
We observe the rapid brightening as a fast increase in intensity over a significant length of the existing fibril: it may cover several hundreds of kilometers in a few seconds.
The intensity increase is mostly uniform over these lengths, and we do not find systematic evidence that the intensity increase grows from the bottom and up.
We find that this behavior is at odds with the description of 
\citet{2007Sci...318.1594K}, 
who remarked about the evolution that the brightening seems to start from the root of the microjet and that the intensity pattern provides an apparent rise velocity faster than 100~\kms\ in the initial phase.
For almost all events in our data set, we cannot identify a clear root or source from where the brightening of the PMJs appears to originate.
Rather, most of the PMJs brighten coherently over considerable length.

We identify three events that appear to have a $>100$~\kms\ rise phase as testified by a steep slope in the space-time diagram: PMJ~I, P, and Q. 
It is questionable that the $\sim220$~\kms\ slope in the space-time diagram for PMJ~I in Fig.~\ref{fig:detailsIJ} can be interpreted as the rise of a single PMJ. This event is unlike the other PMJs in our sample and appears as the fast (and likely temporally unresolved) evolution of a collection of separated blobs. 
PMJ~P in Fig.~\ref{fig:app_detailsNOP} has a $\sim140$~\kms\ slope in the space-time diagram in the later phase of its evolution. 
The accompanying movie gives the distinct impression that there is fast upward propagation along the PMJ throughout its lifetime. 
However, when we zoom in on this rise phase in Fig.~\ref{fig:app_onset}, the rise appears to be more on the order of 25~\kms.
PMJ~Q in Fig.~\ref{fig:app_detailsQ} has a $~125$~\kms\ slope in the space-time diagram and is the only event in our sample that has such fast and apparently resolved rise during the onset of the PMJ. 
Figure~\ref{fig:app_onset} zooms in on the onset of PMJ~Q, and we see a rise of the top of the 90\% level of 370~km in 3~s, which corresponds to an apparent rise velocity of about 120~\kms. 

At least two PMJs in our sample apparently propagate downward (i.e., in the direction toward the umbra) during their onset: PMJs~D and F.
The apparent velocity for PMJ~F is on the order of 220~\kms\ (see Fig.~\ref{fig:onset}), and for PMJ~D, it is 112~\kms\ (see Fig.~\ref{fig:app_onset}).

For the other PMJs, the intensity rise during the onset appears to be coherent over hundreds of kilometers, and there is no clear slope in the space-time diagrams that suggests a rise or downward propagation. 
During the onset, the length over which a PMJ crosses an (arbitrary) intensity threshold of 90\% of PMJ peak intensity can, for example, be 370~km in 1~s (PMJ~A), 200~km in 1~s (B), 600~km in 3~s (D), 622~km in 3~s (E), and 150~km in 1~s (F and G). 
These numbers do not constitute an apparent propagation speed but rather are a measure of the length over which the PMJ intensity is growing rapidly. 
After the fast intensity rise during the onset, the length of the PMJ may continue to grow at both the top and bottom ends of the PMJ. 

This is another notable result from our analysis: the PMJ evolution after the fast onset. 
For about half of the selected events, the top of the PMJ moves with an apparent velocity between 8 and 14~\kms. 
Most of these rise (20 PMJs), but 3 PMJs have a downward-moving top.
For the remaining PMJs (22), there is no significant motion of the top. 

For slightly more than half the PMJs (24), the bottom end moves upward toward the end of the PMJ lifetime. 
For the cases where the upward motion of the bottom end is faster than the top end, the PMJ appears to be shrinking and gives an impression of an outward motion of the PMJ (away from the umbra). See, for example, PMJs A, B, L, and N.
PMJs G and M appear to be small elongated blobs that are moving upward throughout their entire lifetime. 

Seeing might be the reason that we do not observe the fast rise during the PMJ onset, as earlier reported from \Hinode\ observations.
Like for all ground-based observations, seeing distortions and their effect on the analysis are a concern.
We note that the seeing conditions were excellent, the adaptive optics system was performing well, and the image restoration contributed to further improve the reduced data. 
Still, the short-term variability we see in the PMJ evolution can at least partly be attributed to seeing.
As a measure of the seeing quality, Figs.~\ref{fig:detailsAB}--\ref{fig:detailsIJ} (and Figs.~\ref{fig:app_detailsKLM}--\ref{fig:app_detailsQ}) include a panel that shows the contrast variation of the wideband channel. 
Seeing distortions lead to reduced contrast, and some of the contrast dips correlate well with dips in the PMJ intensity. 
A clear example is shown for PMJ~C  in Fig.~\ref{fig:detailsCD}, where a period with very poor seeing after $t=10$ results in a clear dip in the PMJ brightness. 
Another (more subtle) example is PMJ~D in the same figure around $t=5$, or the variation in the peak intensity for PMJ~H (Fig.~\ref{fig:detailsGH}). 
Nevertheless, we consider it unlikely that seeing distortions are the reason that we do not observe a prevalent $>100$~\kms\ apparent rise velocity during the onset phase of the PMJs in our data set.
We are confident that the data are of sufficient quality to reveal this property if it were present, in particular because we observe a fast apparent rise for PMJ~Q and fast downward propagation for PMJs~D and F. 

When we consider the impact of seeing on the apparent evolution of PMJs in our data, we do caution, however, that it would be challenging to distinguish actual (solar) intensity variations from seeing-induced variations in the short-term intensity variability we see during the bright PMJ phase in the light curves of Figs.~\ref{fig:detailsAB}--\ref{fig:detailsIJ}.
Intensity modulations may be expected as a result of magnetohydronynamics (MHD) waves in PMJs, but the effect of seeing on the variability is difficult to quantify. 

\begin{figure*}
\sidecaption
\includegraphics{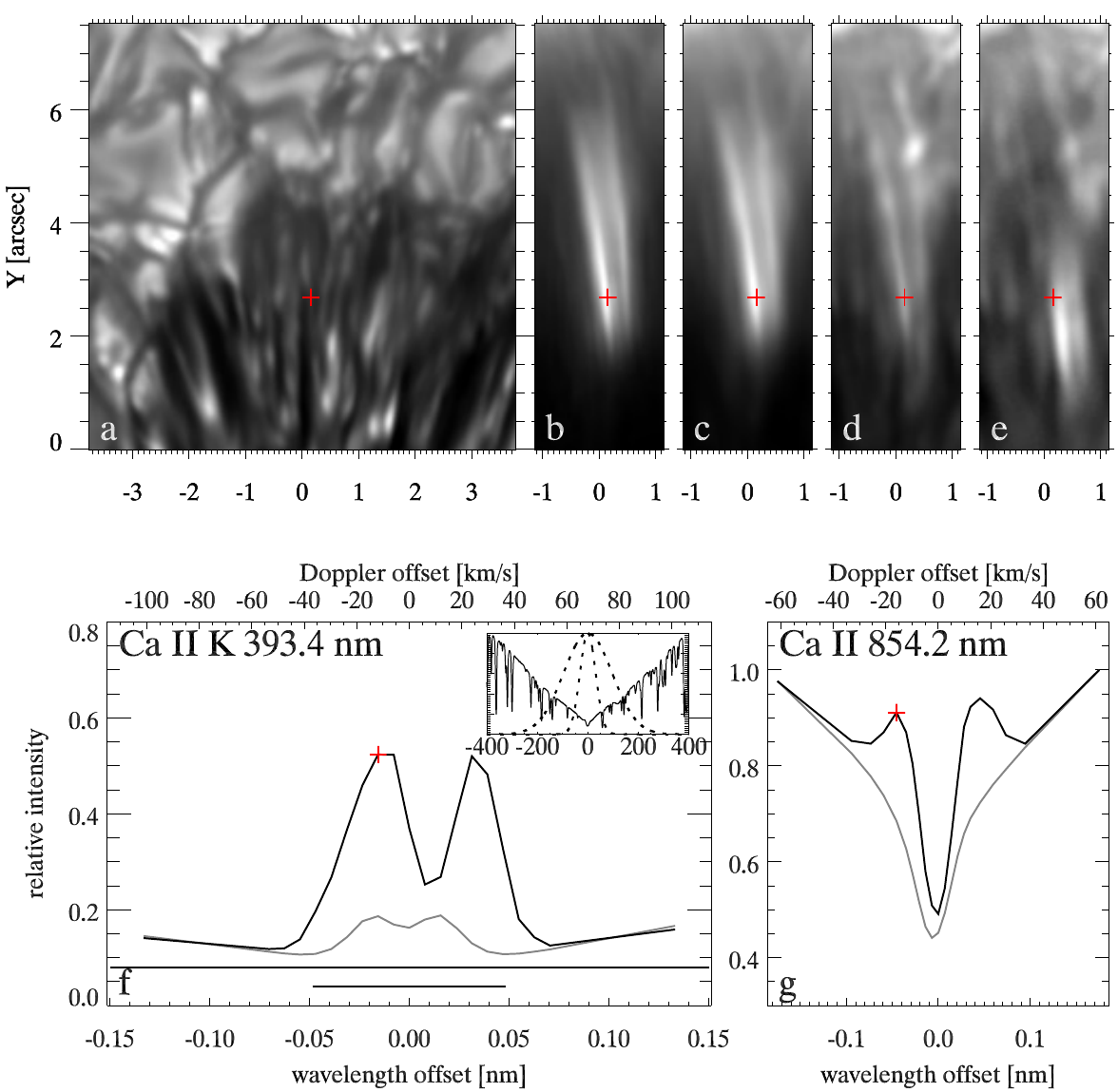}
\caption{%
Examples of spectrally resolved PMJs. Panel a) shows a photospheric CHROMIS wideband reference image (center wavelength 395.0~nm and bandwidth FWHM=1.32~nm). Panel b) shows a CHROMIS \CaK\ image at a Doppler offset of $-12$~\kms. Panel c) shows a CHROMIS \CaK\ image integrated over the full sampled spectral profile, convolved with a Gaussian profile with FWHM=0.1~nm. Panel d) shows a similar integrated CHROMIS \CaK\ image, convolved with a FWHM=0.3~nm Gaussian, and giving enhanced weight to the two outer scan positions. Panel e) shows a CRISP \CaIR\ image at a Doppler offset of $-16$~\kms. The red crosses in the images mark the spatial location in the PMJ for which spectral profiles are shown in panels f) and g). The thick solid lines in panels f) and g) show the \CaK\ and \CaIR\ PMJ spectral profiles, and the red crosses mark the spectral positions of panels b) and e), respectively. The gray lines are reference spectral profiles averaged over a region outside the active region. Panel f) shows horizontal lines of width 0.3 and 0.1~nm as reference to the FWHM bandwidth of the \Hinode\ and SST imaging filters. The inset in panel f) shows a \CaH\ spectral profile from the FTS atlas 
\citep{1999SoPh..184..421N} 
against Doppler velocity offset. The dashed lines are Gaussian profiles with FWHM=0.3 and 0.1~nm.  
}
\label{fig:diag}
\end{figure*}

We note that the appearance of PMJs in imaging data is strongly dependent on the transmission bandpass of the instrument. 
For example, the contrast we find for the PMJs in our sample is considerably higher (average 34\%) than the 10--20\% reported by 
\citet{2007Sci...318.1594K}. 
The \Hinode\ \CaH\ filter has FWHM=0.3~nm, which is equivalent to an offset of $\pm$113~\kms\ in Doppler velocity.
The SST \CaH\ filter has FWHM=0.1~nm or $\pm$~41~\kms\ in Doppler velocity.
This means that in principle, a high-velocity feature with a significant line-of-sight component could be Doppler shifted out of the SST bandpass while it would still be covered in the wings of the \Hinode\ bandpass. 
This effect is not a concern, however, because we only considered PMJs with a sizable linear extension in the plane of the sky, which arguably implies that these PMJs have a limited line-of-sight velocity component.
On the other hand, the wide \Hinode\ filter receives significant contribution from the \CaH\ wings that are formed in the upper photosphere, and faint chromospheric emission from the \CaH\ core may drown in the wing contribution from the lower atmosphere. 
Detection of PMJs in \Hinode\ \CaH\ images indeed requires image enhancement such as unsharp masking or running differences to make a clean separation of the PMJ from the background. 
In the SST data this is not necessary as the PMJs stand out clearly against the penumbral surroundings and the peak intensity is on average higher than the average \CaH\ intensity outside the sunspot.

Figure~\ref{fig:diag} illustrates this effect based on CHROMIS \CaK\ spectral imaging (under the reasonable assumption that PMJs show similar spectral properties in \CaK\ as in \CaH). 
A PMJ stands out clearly in the K$_{2\mathrm{V}}$ peak wavelength in panel b), and is also clearly visible in an image integrated over the full CHROMIS \CaK\ scan accounting for an FWHM=0.1~nm Gaussian profile (panel c).
When convolved with a 0.3~nm wide Gaussian, however, the PMJ is much fainter (panel d; note that the outer wings were linearly extrapolated because the CHROMIS scan only covered out to $\pm$101~\kms).
For reference, panel e) shows an image in the \CaIR\ wing at the wavelength of the emission feature in the blue wing. 
Interestingly, the PMJ signal in \CaIR\ is considerably shorter and offset toward the bottom part as compared to the PMJ in \CaK;\  we attribute this to formation height difference between the two spectral lines. 

The PMJ \CaIR\ spectral profile shown in panel g) shows the typical profile shape with peaks in the blue and red wing. This profile shape was studied in detail by 
\citet{2017A&A...602A..80D} 
and was reported earlier by 
\citet{2013ApJ...779..143R}. 
Panel f) shows the \CaK\ profile from the same location in the PMJ, and we see enhancement of the K$_2$ peaks corresponding to the peaks in \CaIR. 
\citet{2019ApJ...870...88E} 
presented more CHROMIS PMJ \CaK\ profiles. 
These spectral profiles suggest that a 0.1~nm \CaH\ (or K) filter has a sufficiently wide spectral transmission bandwidth to integrate the enhanced central reversal emission peaks, which is typical for PMJs. 
In addition, this more narrow transmission filter has the advantage of receiving only little contribution from the inner wings, which originate from the lower parts of the atmosphere that are unrelated to PMJs. 

After considering seeing and the imaging filter properties, we conclude that the high temporal cadence of our data is the crucial difference that explains why we arrive at a different view of the onset of PMJs. 
The apparent rise velocity of $>100$~\kms\ reported by 
\citet{2007Sci...318.1594K} 
was based on data with a cadence of 8--16~s.
The 250~\kms\ estimate for one penumbral jet by
\citet{2016ApJ...816...92T} 
was based on 5--15~s cadence data.  When we consider that the onset phase is typically shorter than 10~s, it is clear that faster cadence data are required in order to properly resolve the dynamics during the time in which the PMJ rapidly brightens. 

\citet{2017ApJ...835L..19S} 
studied the relation between subarcsecond bright dots in \IRIS\ 1400~\AA\ slit-jaw images and PMJs observed in an 1.58~s cadence \Hinode\ \CaH\ time series. 
Their work does not focus on the speed of the \CaH\ PMJs, but they noted that the PMJs appear as sudden brightenings in their space-time diagrams and that it was very difficult to determine their propagation direction and speed. They further remarked that it is unclear whether the PMJs move radially outward from the sunspot. 
This description agrees well with our results. 

We conclude that mass upflows can play only a limited role in the onset phase of PMJs. 
The intensity light curve (particularly considering the time and length scale) we observe during the rapid rise cannot be reconciled with a pure mass upflow. 
This was already clear from the spectral inversions by
\citet{2019ApJ...870...88E} , 
   who showed that PMJs do not harbor higher line-of-sight velocities than their surroundings and the inferred velocities did not exceed $\pm4$~\kms. 
Earlier analyses of \CaIR\ line profiles by 
\citet{2013ApJ...779..143R} 
and \citet{2017A&A...602A..80D} 
did not reveal any high-velocity spectral signatures. 
Furthermore, the latter did not find a viewing-angle correlation in the line offset of the \CaIR\ profile features, which might be expected if there were a systematic mass flow in PMJs. 

This disparity between inferred Doppler velocity and apparent dynamical velocity recalls observed characteristics of type II spicules, although less extreme than for PMJs:  
Type II spicules display apparent rise velocities at the limb of 30--150~\kms\
\citep{2007PASJ...59S.655D, 
2012ApJ...759...18P} 
, and their transition region (TR) counterparts (also known as network jets) appear to move with velocities 80--300~\kms\ on the disk
\citep{2014Sci...346A.315T, 
2016SoPh..291.1129N}. 
Inferred Doppler velocities, however, are typically in the range 20--50~\kms\ 
\citep{2009ApJ...705..272R, 
2012ApJ...752..108S} 
or 50--70~\kms\ in TR diagnostics 
\citep{2015ApJ...799L...3R}. 

\citet{2017ApJ...849L...7D} 
used advanced radiative MHD simulations to explain the disparity between Doppler and apparent velocities in spicules. 
In these simulations, relatively cool spicular material undergoes rapid heating events through the dissipation of currents by ambipolar diffusion. 
These currents propagate at Alfv\'enic speeds (well above 100~\kms)  through preexisting spicules and cause rapid heating that results in a sudden brightening and associated fast apparent motion in chromospheric and TR diagnostics. 
They concluded that the observed fast apparent motions in spicules are a signature of a heating front that propagates at much higher velocity than the mass flow in spicules. 
More support for this scenario was recently found by 
\citet{2018ApJ...857...73C} , 
who studied a type II spicule in co-temporal VAULT2.0 \Lyalpha\ and IRIS TR observations. 
A spicule was found to exist at low \Lyalpha\ intensity already 2 min before it appeared as a fast 300~\kms\ network jet in IRIS \ion{C}{ii}~1330\AA\ and \ion{Si}{iv}~1400\AA. 
At the time the spicule appeared as an IRIS TR network jet, the spicule also brightened in \Lyalpha.
This favors the interpretation that the cool preexisting spicule was heated by rapid current dissipation within the spicule so that it became visible in the TR diagnostics. 
A multitemperature scenario for spicules was also inferred from combined observations obtained with the SST \Halpha\ and the Atmospheric Imaging Assembly onboard the Solar Dynamcs Observatory (SDO/AIA) of the 304\AA\ and 171\AA\ lines by 
\citet{2016ApJ...820..124H}. 

\citet{2019ApJ...870...88E} 
speculated that this spicule scenario of a heating front also holds for PMJ formation.
Several aspects of the observations we presented here further support this view. 
First, we find evidence that PMJs appear as brightenings of preexisting fibrils.
Second, the fast brightenings over the spatial scales we observe cannot be explained by mass motions. 
If the PMJs brighten as a result of a heating front, the associated current must propagate at extremely high speed because we find no systematic evidence for propagation speeds in our data. 
We note, however, that the Alfv\'en speed in the simulated spicules of 
\citet{2017ApJ...849L...7D} 
ranges between 150--450~\kms. 
The Alfv\'en speed in PMJs may be even higher because the magnetic field strength in the penumbra is higher than in the network areas where spicules originate. 
We further note that the synthetic \ion{Si}{iv} emission in the space-time diagrams of 
\citet{2017ApJ...849L...7D} 
is nearly vertical for the simulated spicules, not unlike the PMJ space-time diagrams we presented here. 
The apparent motion of the PMJ top that occurs at more sonic speeds ($<14$~\kms) might be due to expansion after the rapid heating phase. The downward motion that we observe for a few PMJs may be contraction due to cooling.
In any case, advanced models that correctly model the complex and strong magnetic field environment of the penumbra are required to further advance our understanding of PMJs and the penumbral chromosphere.
Our observations provide stringent constraints to such modeling efforts. 

A significant number of the PMJs in our sample (14 PMJs, or 31\%) appear to be split into two parallel and coevolving structures during the advanced stages of their dynamical evolution. 
When we assume that there is no preferred splitting direction, then splitting may be even more common because PMJs that split along a direction with small inclination angle to the line of sight will not be resolved into separated structures.
\citet{2008ApJ...686.1404R} 
found splitting into double structures to be common in features that they referred to as chromospheric transients. 
They studied different \Hinode\ data sets of a sunspot during its passage over the solar disk. 
They distinguished between microjets and a type of elongated chromospheric brightening that is longer lived, has longer length, and displays clear sideways motion. The latter are interpreted as a result of a bow shock after reconnection in neighboring penumbral filaments. Splitting was found to be common in this type of transient and explained in the context of the bow-shock scenario.
We note that the splitting PMJs in our sample are shorter, have shorter lifetimes, and display less sideways motion compared to the characteristics that 
\citet{2008ApJ...686.1404R} 
reported. 

For the sake of reference, we have measured FWHM lengths and widths of the PMJs in our sample. 
They are slightly shorter and narrower than the measurements by
\citet{2019ApJ...870...88E} 
, who based their measurements on locating the boundaries of \CaIR\ PMJ spectral profiles.
This naturally produces higher values than an FWHM measurement. 
This might also explain why we found shorter lengths than 
\citet{2007Sci...318.1594K}, 
while the narrower width can also be attributed to the higher spatial resolution of the SST compared to \Hinode.

\section{Summary and conclusions}
\label{sec:conclusion}

We have studied the dynamical evolution of PMJs in high spatial and temporal (1~s cadence) resolution \CaH\ filtergram time series of a sunspot observed on two consecutive days. 
With the narrow transmission passband of the imaging filter (FWHM 0.1~nm), PMJs stand out as high-contrast linear features that do not require image enhancement techniques to facilitate identification. 
We selected 45 PMJs that had consistent good seeing conditions throughout their lifetimes for further detailed analysis. 
Our results can be summarized as follows:
\begin{enumerate}
    \item The PMJs appear to be the rapid brightening of a preexisting (faint) fibril. After the PMJ has faded again, a faint remaining fibril can often be observed.
    \item The rapid brightening is the fast increase in intensity over a significant length of the existing fibril. During the onset phase, when the intensity rises at its fastest, the increase can be on the order 10--40\% of the peak PMJ intensity in less than 10~s. During the onset phase, the length over which the PMJ rises above a threshold of 90\% can be between 150--370~km in 1~s.
    \item For almost no event in our data set can we identify a clear root or source from where the brightening of the PMJs would appear to originate.
    \item After the fast intensity rise, about half of the PMJs have a top that moves with an apparent velocity of between 5
    and 14~\kms. Twenty PMJs have a rising top, 3 have a downward-moving top. The remaining 20 PMJs show no significant motion of the top. 
    \item Toward the end of the lifetime of little more then half of the PMJs (24), the bottom end of the PMJ rises upward. This can result in an apparent shrinking of the PMJ, and in some cases (if the PMJ top rises as well), the whole PMJ appears to be moving away from the umbra. 
    \item The average intensity increase of the PMJs is 34\% (ranging between 15\% and 75\%). Compared to the quieter surroundings of the sunspot, the PMJs reach on average 1.5 times the quiet intensity (ranging between a factor 0.5 and 2.8).
    \item The average FWHM length of the PMJs at the time of maximum intensity is 802~km (ranging between 200 and 1630~km), and the average FWHM width is 179~km (ranging between 85 and 415~km).
    \item For about one-third of the PMJs (14) we observe a splitting into two parallel and coevolving linear features during the later phases of the PMJ lifetimes. 
\end{enumerate}

In conclusion, the most stringent constraints on PMJ modeling efforts are placed by the finding that PMJs show fast brightening of preexisting fibrils over lengths of hundreds of kilometers. 
The fact that at low temporal resolution the onset phase can be mistaken for mass motion underlines the importance of high temporal resolution when the dynamic solar atmosphere is studied.
This is a message that should not be forgotten in an era when high spatial resolution and photon-gathering power is driving the development of instrumentation for the next generation of 4m class solar telescopes.

\begin{acknowledgements}
The Swedish 1-m Solar Telescope is operated on the island of La Palma
by the Institute for Solar Physics of Stockholm University in the
Spanish Observatorio del Roque de los Muchachos of the Instituto de
Astrof{\'\i}sica de Canarias.
The Institute for Solar Physics is supported by a grant for research infrastructures of national importance from the Swedish Research Council (registration number 2017-00625).
This research is supported by the Research Council of Norway, project number 250810, and through its Centres of Excellence scheme, project number 262622.
This study benefited from discussions during the workshop ``Studying magnetic-field-regulated heating in the solar chromosphere'' (team 399) at the International Space Science Institute (ISSI) in Switzerland.
We made much use of NASA's Astrophysics Data System Bibliographic
Services.

\end{acknowledgements}


\begin{thebibliography}{35}
\expandafter\ifx\csname natexlab\endcsname\relax\def\natexlab#1{#1}\fi

\bibitem[{{Borrero} \& {Ichimoto}(2011)}]{2011LRSP....8....4B}
{Borrero}, J.~M. \& {Ichimoto}, K. 2011, Living Reviews in Solar Physics, 8, 4
  \csname 2011LRSP....8....4Blink\endcsname~\csname
  2011LRSP....8....4Bnote\endcsname

\bibitem[{{Chintzoglou} {et~al.}(2018){Chintzoglou}, {De Pontieu},
  {Mart{\'{\i}}nez-Sykora}, {Pereira}, {Vourlidas}, \& {Tun
  Beltran}}]{2018ApJ...857...73C}
{Chintzoglou}, G., {De Pontieu}, B., {Mart{\'{\i}}nez-Sykora}, J., {et~al.}
  2018, \apj, 857, 73 \csname 2018ApJ...857...73Clink\endcsname~\csname
  2018ApJ...857...73Cnote\endcsname

\bibitem[{{de la Cruz Rodr{\'{\i}}guez} {et~al.}(2015){de la Cruz
  Rodr{\'{\i}}guez}, {L{\"o}fdahl}, {S{\"u}tterlin}, {Hillberg}, \& {Rouppe van
  der Voort}}]{2015A&A...573A..40D}
{de la Cruz Rodr{\'{\i}}guez}, J., {L{\"o}fdahl}, M.~G., {S{\"u}tterlin}, P.,
  {Hillberg}, T., \& {Rouppe van der Voort}, L. 2015, \aap, 573, A40 \csname
  2015A&A...573A..40Dlink\endcsname~\csname 2015A&A...573A..40Dnote\endcsname

\bibitem[{{De Pontieu} {et~al.}(2017){De Pontieu}, {Mart{\'{\i}}nez-Sykora}, \&
  {Chintzoglou}}]{2017ApJ...849L...7D}
{De Pontieu}, B., {Mart{\'{\i}}nez-Sykora}, J., \& {Chintzoglou}, G. 2017,
  \apjl, 849, L7 \csname 2017ApJ...849L...7Dlink\endcsname~\csname
  2017ApJ...849L...7Dnote\endcsname

\bibitem[{{De Pontieu} {et~al.}(2007){De Pontieu}, {McIntosh}, {Hansteen},
  {Carlsson}, {Schrijver}, {Tarbell}, {Title}, {Shine}, {Suematsu}, {Tsuneta},
  {Katsukawa}, {Ichimoto}, {Shimizu}, \& {Nagata}}]{2007PASJ...59S.655D}
{De Pontieu}, B., {McIntosh}, S., {Hansteen}, V.~H., {et~al.} 2007, \pasj, 59,
  S655 \csname 2007PASJ...59S.655Dlink\endcsname~\csname
  2007PASJ...59S.655Dnote\endcsname

\bibitem[{{Drews} \& {Rouppe van der Voort}(2017)}]{2017A&A...602A..80D}
{Drews}, A. \& {Rouppe van der Voort}, L. 2017, \aap, 602, A80 \csname
  2017A&A...602A..80Dlink\endcsname~\csname 2017A&A...602A..80Dnote\endcsname

\bibitem[{{Esteban Pozuelo} {et~al.}(2019){Esteban Pozuelo}, {de la Cruz
  Rodr{\'{\i}}guez}, {Drews}, {Rouppe van der Voort}, {Scharmer}, \&
  {Carlsson}}]{2019ApJ...870...88E}
{Esteban Pozuelo}, S., {de la Cruz Rodr{\'{\i}}guez}, J., {Drews}, A., {et~al.}
  2019, \apj, 870, 88 \csname 2019ApJ...870...88Elink\endcsname~\csname
  2019ApJ...870...88Enote\endcsname

\bibitem[{{Henriques} {et~al.}(2016){Henriques}, {Kuridze}, {Mathioudakis}, \&
  {Keenan}}]{2016ApJ...820..124H}
{Henriques}, V.~M.~J., {Kuridze}, D., {Mathioudakis}, M., \& {Keenan}, F.~P.
  2016, \apj, 820, 124 \csname 2016ApJ...820..124Hlink\endcsname~\csname
  2016ApJ...820..124Hnote\endcsname

\bibitem[{{Katsukawa} {et~al.}(2007){Katsukawa}, {Berger}, {Ichimoto}, {Lites},
  {Nagata}, {Shimizu}, {Shine}, {Suematsu}, {Tarbell}, {Title}, \&
  {Tsuneta}}]{2007Sci...318.1594K}
{Katsukawa}, Y., {Berger}, T.~E., {Ichimoto}, K., {et~al.} 2007, Science, 318,
  1594 \csname 2007Sci...318.1594Klink\endcsname~\csname
  2007Sci...318.1594Knote\endcsname

\bibitem[{{Katsukawa} \& {Jur{\v c}{\'a}k}(2010)}]{2010A&A...524A..20K}
{Katsukawa}, Y. \& {Jur{\v c}{\'a}k}, J. 2010, \aap, 524, A20 \csname
  2010A&A...524A..20Klink\endcsname~\csname 2010A&A...524A..20Knote\endcsname

\bibitem[{{Kosugi} {et~al.}(2007){Kosugi}, {Matsuzaki}, {Sakao}, {Shimizu},
  {Sone}, {Tachikawa}, {Hashimoto}, {Minesugi}, {Ohnishi}, {Yamada}, {Tsuneta},
  {Hara}, {Ichimoto}, {Suematsu}, {Shimojo}, {Watanabe}, {Shimada}, {Davis},
  {Hill}, {Owens}, {Title}, {Culhane}, {Harra}, {Doschek}, \&
  {Golub}}]{2007SoPh..243....3K}
{Kosugi}, T., {Matsuzaki}, K., {Sakao}, T., {et~al.} 2007, \solphys, 243, 3
  \csname 2007SoPh..243....3Klink\endcsname~\csname
  2007SoPh..243....3Knote\endcsname

\bibitem[{{L{\"o}fdahl} {et~al.}(2011){L{\"o}fdahl}, {Henriques}, \&
  {Kiselman}}]{2011A&A...533A..82L}
{L{\"o}fdahl}, M.~G., {Henriques}, V.~M.~J., \& {Kiselman}, D. 2011, \aap, 533,
  A82 \csname 2011A&A...533A..82Llink\endcsname~\csname
  2011A&A...533A..82Lnote\endcsname

\bibitem[{{L{\"o}fdahl} {et~al.}(2018){L{\"o}fdahl}, {Hillberg}, {de la Cruz
  Rodriguez}, {Vissers}, {Scharmer}, {Hagfors Haugan}, \&
  {Fredvik}}]{2018arXiv180403030L}
{L{\"o}fdahl}, M.~G., {Hillberg}, T., {de la Cruz Rodriguez}, J., {et~al.}
  2018, ArXiv e-prints, 1804.03030 \csname
  2018arXiv180403030Llink\endcsname~\csname 2018arXiv180403030Lnote\endcsname

\bibitem[{{Nakamura} {et~al.}(2012){Nakamura}, {Shibata}, \&
  {Isobe}}]{2012ApJ...761...87N}
{Nakamura}, N., {Shibata}, K., \& {Isobe}, H. 2012, \apj, 761, 87 \csname
  2012ApJ...761...87Nlink\endcsname~\csname 2012ApJ...761...87Nnote\endcsname

\bibitem[{{Narang} {et~al.}(2016){Narang}, {Arbacher}, {Tian}, {Banerjee},
  {Cranmer}, {DeLuca}, \& {McKillop}}]{2016SoPh..291.1129N}
{Narang}, N., {Arbacher}, R.~T., {Tian}, H., {et~al.} 2016, \solphys, 291, 1129
  \csname 2016SoPh..291.1129Nlink\endcsname~\csname
  2016SoPh..291.1129Nnote\endcsname

\bibitem[{{Neckel}(1999)}]{1999SoPh..184..421N}
{Neckel}, H. 1999, \solphys, 184, 421 \csname
  1999SoPh..184..421Nlink\endcsname~\csname 1999SoPh..184..421Nnote\endcsname

\bibitem[{{Pereira} {et~al.}(2012){Pereira}, {De Pontieu}, \&
  {Carlsson}}]{2012ApJ...759...18P}
{Pereira}, T.~M.~D., {De Pontieu}, B., \& {Carlsson}, M. 2012, \apj, 759, 18
  \csname 2012ApJ...759...18Plink\endcsname~\csname
  2012ApJ...759...18Pnote\endcsname

\bibitem[{{Reardon} {et~al.}(2013){Reardon}, {Tritschler}, \&
  {Katsukawa}}]{2013ApJ...779..143R}
{Reardon}, K., {Tritschler}, A., \& {Katsukawa}, Y. 2013, \apj, 779, 143
  \csname 2013ApJ...779..143Rlink\endcsname~\csname
  2013ApJ...779..143Rnote\endcsname

\bibitem[{{Rouppe van der Voort} {et~al.}(2015){Rouppe van der Voort}, {De
  Pontieu}, {Pereira}, {Carlsson}, \& {Hansteen}}]{2015ApJ...799L...3R}
{Rouppe van der Voort}, L., {De Pontieu}, B., {Pereira}, T.~M.~D., {Carlsson},
  M., \& {Hansteen}, V. 2015, \apjl, 799, L3 \csname
  2015ApJ...799L...3Rlink\endcsname~\csname 2015ApJ...799L...3Rnote\endcsname

\bibitem[{{Rouppe van der Voort} {et~al.}(2017){Rouppe van der Voort}, {De
  Pontieu}, {Scharmer}, {de la Cruz Rodr{\'{\i}}guez},
  {Mart{\'{\i}}nez-Sykora}, {N{\'o}brega-Siverio}, {Guo}, {Jafarzadeh},
  {Pereira}, {Hansteen}, {Carlsson}, \& {Vissers}}]{2017ApJ...851L...6R}
{Rouppe van der Voort}, L., {De Pontieu}, B., {Scharmer}, G.~B., {et~al.} 2017,
  \apjl, 851, L6 \csname 2017ApJ...851L...6Rlink\endcsname~\csname
  2017ApJ...851L...6Rnote\endcsname

\bibitem[{{Rouppe van der Voort} {et~al.}(2009){Rouppe van der Voort},
  {Leenaarts}, {de Pontieu}, {Carlsson}, \& {Vissers}}]{2009ApJ...705..272R}
{Rouppe van der Voort}, L., {Leenaarts}, J., {de Pontieu}, B., {Carlsson}, M.,
  \& {Vissers}, G. 2009, \apj, 705, 272 \csname
  2009ApJ...705..272Rlink\endcsname~\csname 2009ApJ...705..272Rnote\endcsname

\bibitem[{{Ryutova} {et~al.}(2008){Ryutova}, {Berger}, {Frank}, \&
  {Title}}]{2008ApJ...686.1404R}
{Ryutova}, M., {Berger}, T., {Frank}, Z., \& {Title}, A. 2008, \apj, 686, 1404
  \csname 2008ApJ...686.1404Rlink\endcsname~\csname
  2008ApJ...686.1404Rnote\endcsname

\bibitem[{{Samanta} {et~al.}(2017){Samanta}, {Tian}, {Banerjee}, \&
  {Schanche}}]{2017ApJ...835L..19S}
{Samanta}, T., {Tian}, H., {Banerjee}, D., \& {Schanche}, N. 2017, \apjl, 835,
  L19 \csname 2017ApJ...835L..19Slink\endcsname~\csname
  2017ApJ...835L..19Snote\endcsname

\bibitem[{{Scharmer} {et~al.}(2003{\natexlab{a}}){Scharmer}, {Bjelksj{\"o}},
  {Korhonen}, {Lindberg}, \& {Petterson}}]{2003SPIE.4853..341S}
{Scharmer}, G.~B., {Bjelksj{\"o}}, K., {Korhonen}, T.~K., {Lindberg}, B., \&
  {Petterson}, B. 2003{\natexlab{a}}, in \procspie, Vol. 4853, Innovative
  Telescopes and Instrumentation for Solar Astrophysics, ed. S.~L. {Keil} \&
  S.~V. {Avakyan}, 341--350 \csname 2003SPIE.4853..341Slink\endcsname~\csname
  2003SPIE.4853..341Snote\endcsname

\bibitem[{{Scharmer} {et~al.}(2003{\natexlab{b}}){Scharmer}, {Dettori},
  {L{\"o}fdahl}, \& {Shand}}]{2003SPIE.4853..370S}
{Scharmer}, G.~B., {Dettori}, P.~M., {L{\"o}fdahl}, M.~G., \& {Shand}, M.
  2003{\natexlab{b}}, in \procspie, Vol. 4853, Innovative Telescopes and
  Instrumentation for Solar Astrophysics, ed. S.~L. {Keil} \& S.~V. {Avakyan},
  370--380 \csname 2003SPIE.4853..370Slink\endcsname~\csname
  2003SPIE.4853..370Snote\endcsname

\bibitem[{{Scharmer} {et~al.}(2008){Scharmer}, {Narayan}, {Hillberg}, {de la
  Cruz Rodriguez}, {L{\"o}fdahl}, {Kiselman}, {S{\"u}tterlin}, {van Noort}, \&
  {Lagg}}]{2008ApJ...689L..69S}
{Scharmer}, G.~B., {Narayan}, G., {Hillberg}, T., {et~al.} 2008, \apjl, 689,
  L69 \csname 2008ApJ...689L..69Slink\endcsname~\csname
  2008ApJ...689L..69Snote\endcsname

\bibitem[{{Sekse} {et~al.}(2012){Sekse}, {Rouppe van der Voort}, \& {De
  Pontieu}}]{2012ApJ...752..108S}
{Sekse}, D.~H., {Rouppe van der Voort}, L., \& {De Pontieu}, B. 2012, \apj,
  752, 108 \csname 2012ApJ...752..108Slink\endcsname~\csname
  2012ApJ...752..108Snote\endcsname

\bibitem[{{Shine} {et~al.}(1994){Shine}, {Title}, {Tarbell}, {Smith}, {Frank},
  \& {Scharmer}}]{1994ApJ...430..413S}
{Shine}, R.~A., {Title}, A.~M., {Tarbell}, T.~D., {et~al.} 1994, \apj, 430, 413
  \csname 1994ApJ...430..413Slink\endcsname~\csname
  1994ApJ...430..413Snote\endcsname

\bibitem[{{Tian} {et~al.}(2014){Tian}, {DeLuca}, {Cranmer}, {De Pontieu},
  {Peter}, {Mart{\'{\i}}nez-Sykora}, {Golub}, {McKillop}, {Reeves}, {Miralles},
  {McCauley}, {Saar}, {Testa}, {Weber}, {Murphy}, {Lemen}, {Title}, {Boerner},
  {Hurlburt}, {Tarbell}, {Wuelser}, {Kleint}, {Kankelborg}, {Jaeggli},
  {Carlsson}, {Hansteen}, \& {McIntosh}}]{2014Sci...346A.315T}
{Tian}, H., {DeLuca}, E.~E., {Cranmer}, S.~R., {et~al.} 2014, Science, 346,
  1255711 \csname 2014Sci...346A.315Tlink\endcsname~\csname
  2014Sci...346A.315Tnote\endcsname

\bibitem[{{Tiwari}(2017)}]{2017arXiv171207174T}
{Tiwari}, S.~K. 2017, ArXiv e-prints, 1712.07174 \csname
  2017arXiv171207174Tlink\endcsname~\csname 2017arXiv171207174Tnote\endcsname

\bibitem[{{Tiwari} {et~al.}(2016){Tiwari}, {Moore}, {Winebarger}, \&
  {Alpert}}]{2016ApJ...816...92T}
{Tiwari}, S.~K., {Moore}, R.~L., {Winebarger}, A.~R., \& {Alpert}, S.~E. 2016,
  \apj, 816, 92 \csname 2016ApJ...816...92Tlink\endcsname~\csname
  2016ApJ...816...92Tnote\endcsname

\bibitem[{{Tsuneta} {et~al.}(2008){Tsuneta}, {Ichimoto}, {Katsukawa}, {Nagata},
  {Otsubo}, {Shimizu}, {Suematsu}, {Nakagiri}, {Noguchi}, {Tarbell}, {Title},
  {Shine}, {Rosenberg}, {Hoffmann}, {Jurcevich}, {Kushner}, {Levay}, {Lites},
  {Elmore}, {Matsushita}, {Kawaguchi}, {Saito}, {Mikami}, {Hill}, \&
  {Owens}}]{2008SoPh..249..167T}
{Tsuneta}, S., {Ichimoto}, K., {Katsukawa}, Y., {et~al.} 2008, \solphys, 249,
  167 \csname 2008SoPh..249..167Tlink\endcsname~\csname
  2008SoPh..249..167Tnote\endcsname

\bibitem[{{van Noort} {et~al.}(2005){van Noort}, {Rouppe van der Voort}, \&
  {L{\"o}fdahl}}]{2005SoPh..228..191V}
{van Noort}, M., {Rouppe van der Voort}, L., \& {L{\"o}fdahl}, M.~G. 2005,
  \solphys, 228, 191 \csname 2005SoPh..228..191Vlink\endcsname~\csname
  2005SoPh..228..191Vnote\endcsname

\bibitem[{{Vissers} \& {Rouppe van der Voort}(2012)}]{2012ApJ...750...22V}
{Vissers}, G. \& {Rouppe van der Voort}, L. 2012, \apj, 750, 22 \csname
  2012ApJ...750...22Vlink\endcsname~\csname 2012ApJ...750...22Vnote\endcsname

\bibitem[{{Vissers} {et~al.}(2015){Vissers}, {Rouppe van der Voort}, \&
  {Carlsson}}]{2015ApJ...811L..33V}
{Vissers}, G.~J.~M., {Rouppe van der Voort}, L.~H.~M., \& {Carlsson}, M. 2015,
  \apjl, 811, L33 \csname 2015ApJ...811L..33Vlink\endcsname~\csname
  2015ApJ...811L..33Vnote\endcsname

\end{thebibliography}

\begin{appendix}
\section{Additional PMJ examples}
\label{app:examples}

In this appendix we provide more figures with details of the evolution of selected PMJs. 


\begin{figure*}[!ht]
\includegraphics[width=\textwidth]{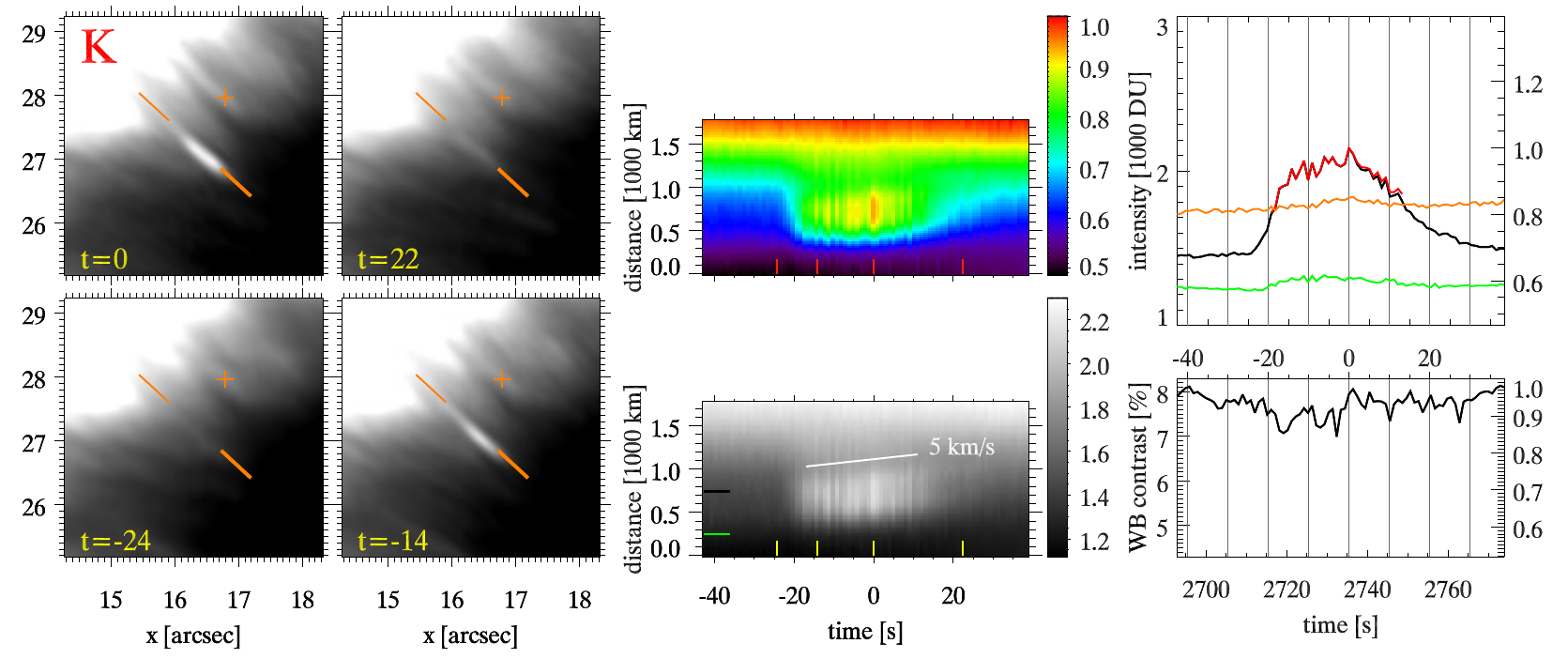} \\
\includegraphics[width=\textwidth]{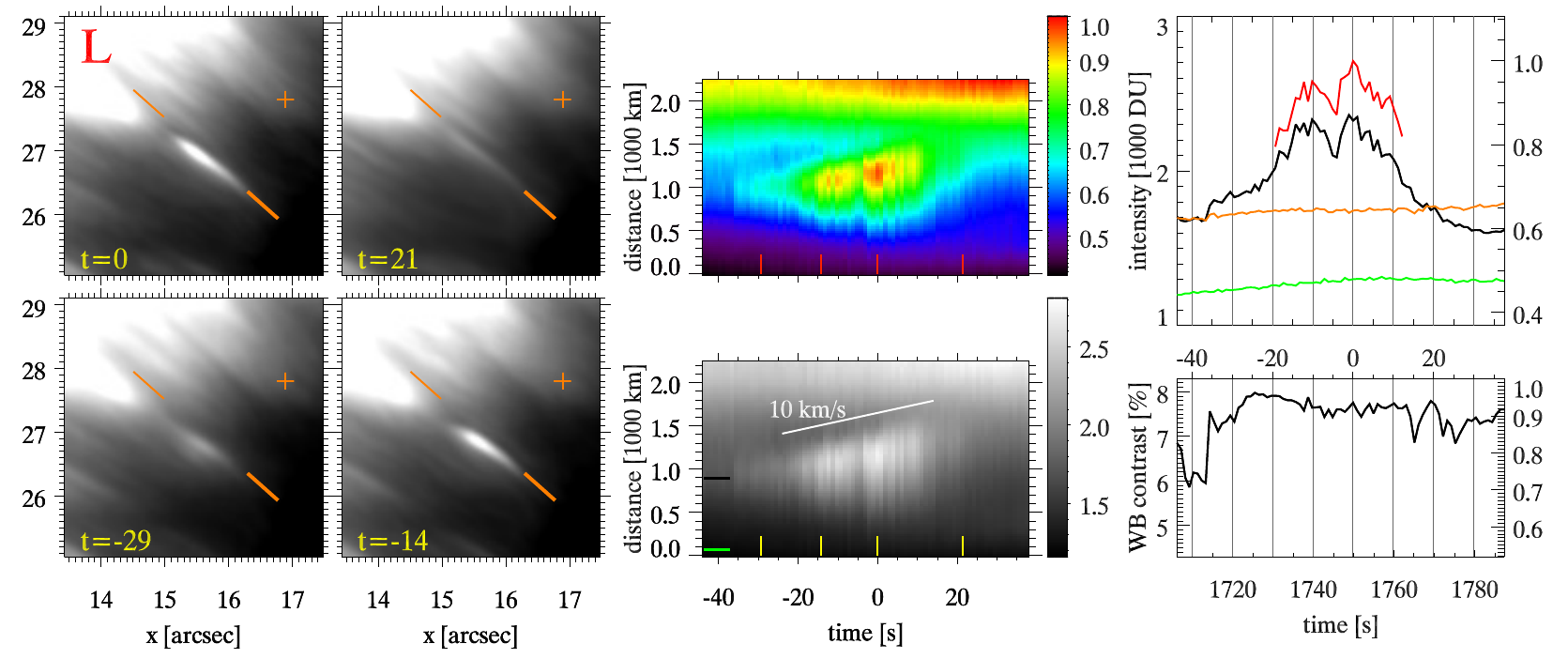} \\
\includegraphics[width=\textwidth]{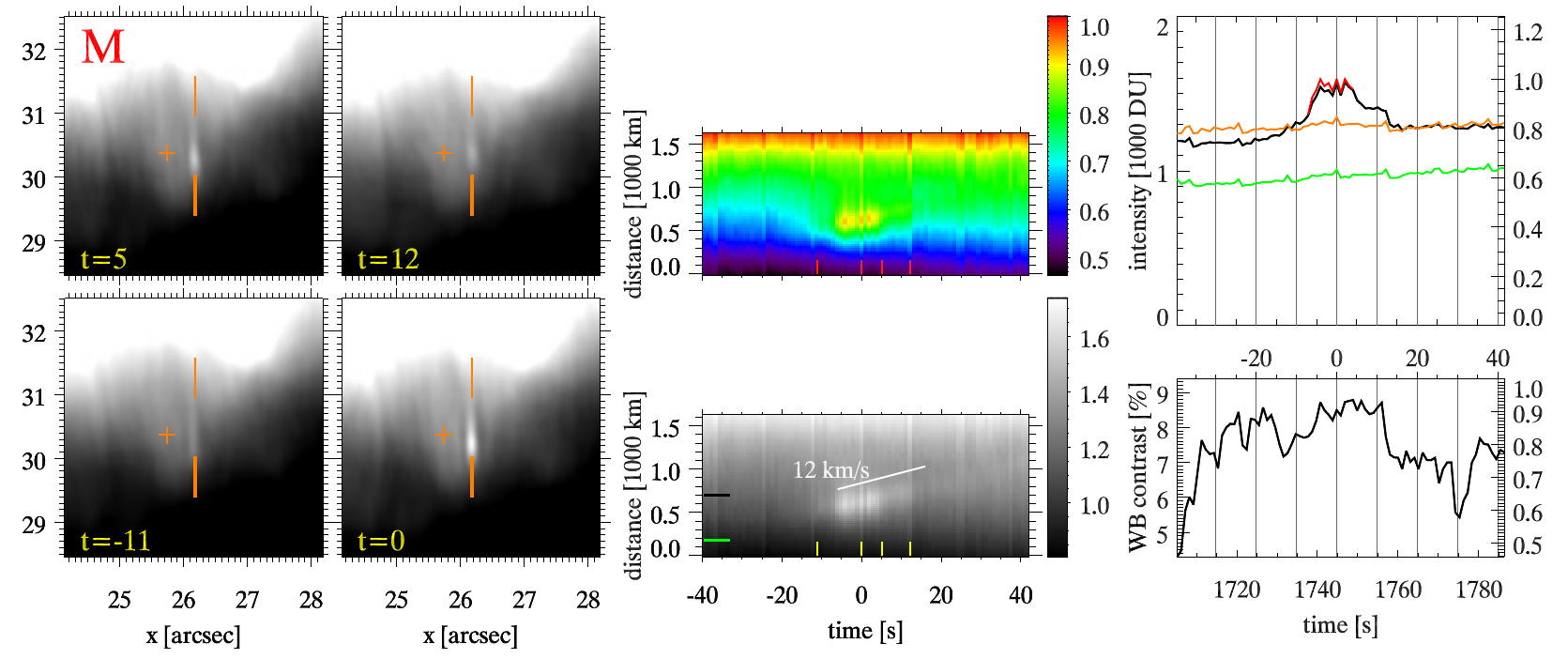}
\caption{\label{fig:app_detailsKLM}%
Details of the evolution of PMJs.
The format of this figure is the same as that of Fig.~\ref{fig:detailsAB}. Animations of this figure are available \url{http://folk.uio.no/rouppe/pmj_highcadence/}.
}
\end{figure*}

\begin{figure*}[!ht]
\includegraphics[width=\textwidth]{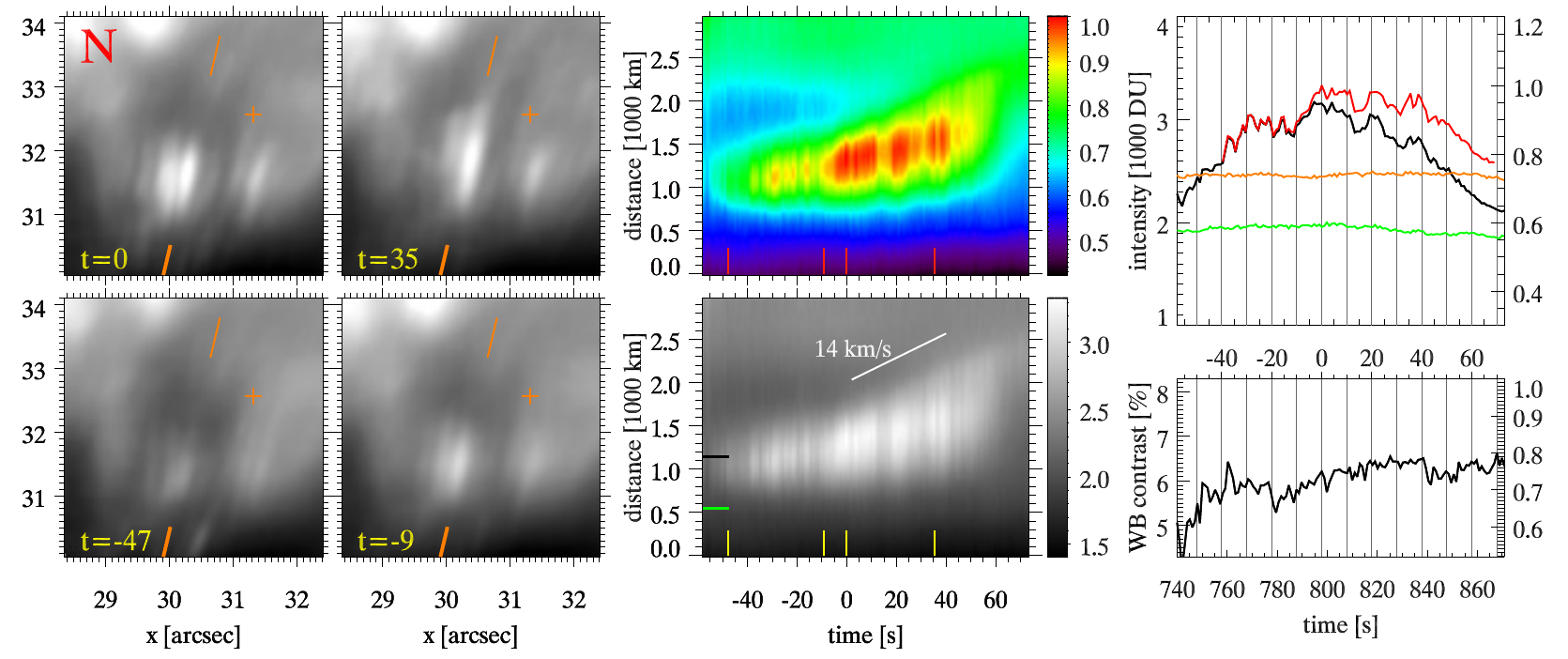} \\
\includegraphics[width=\textwidth]{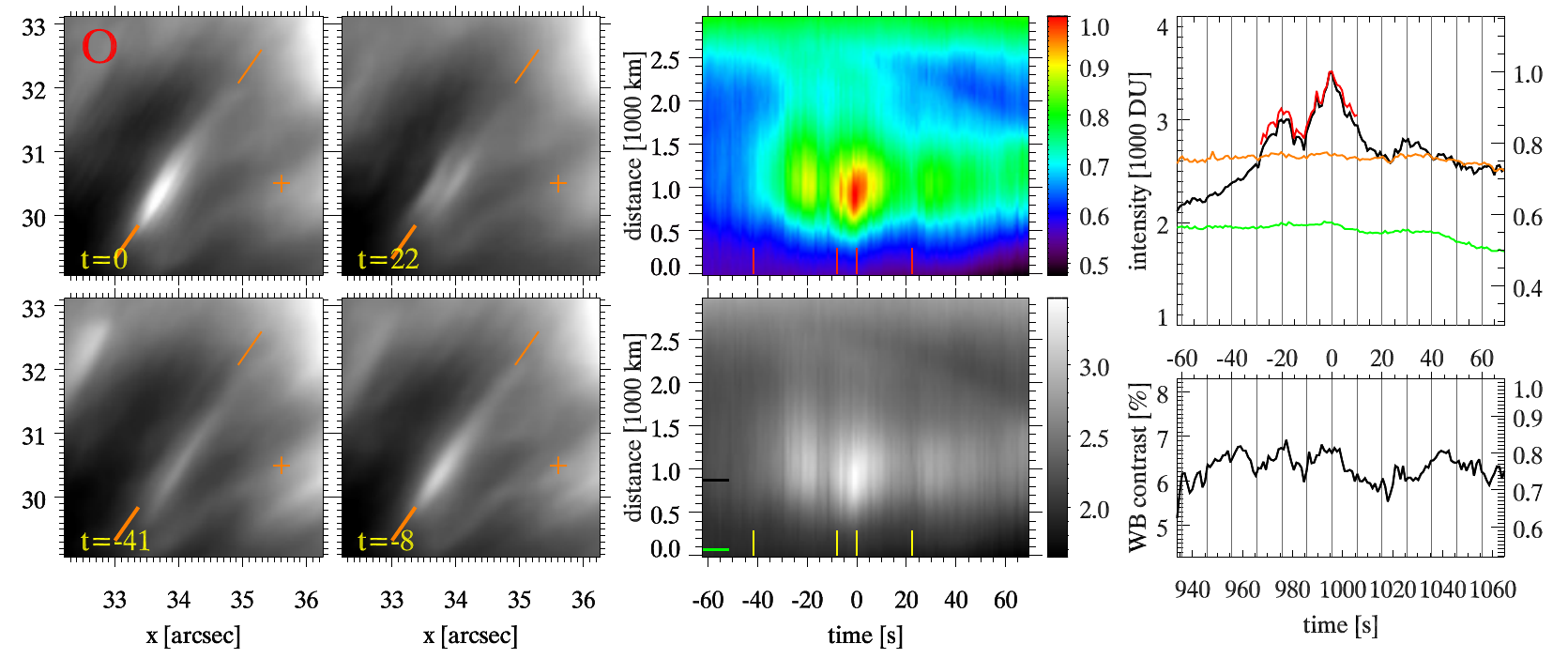} \\
\includegraphics[width=\textwidth]{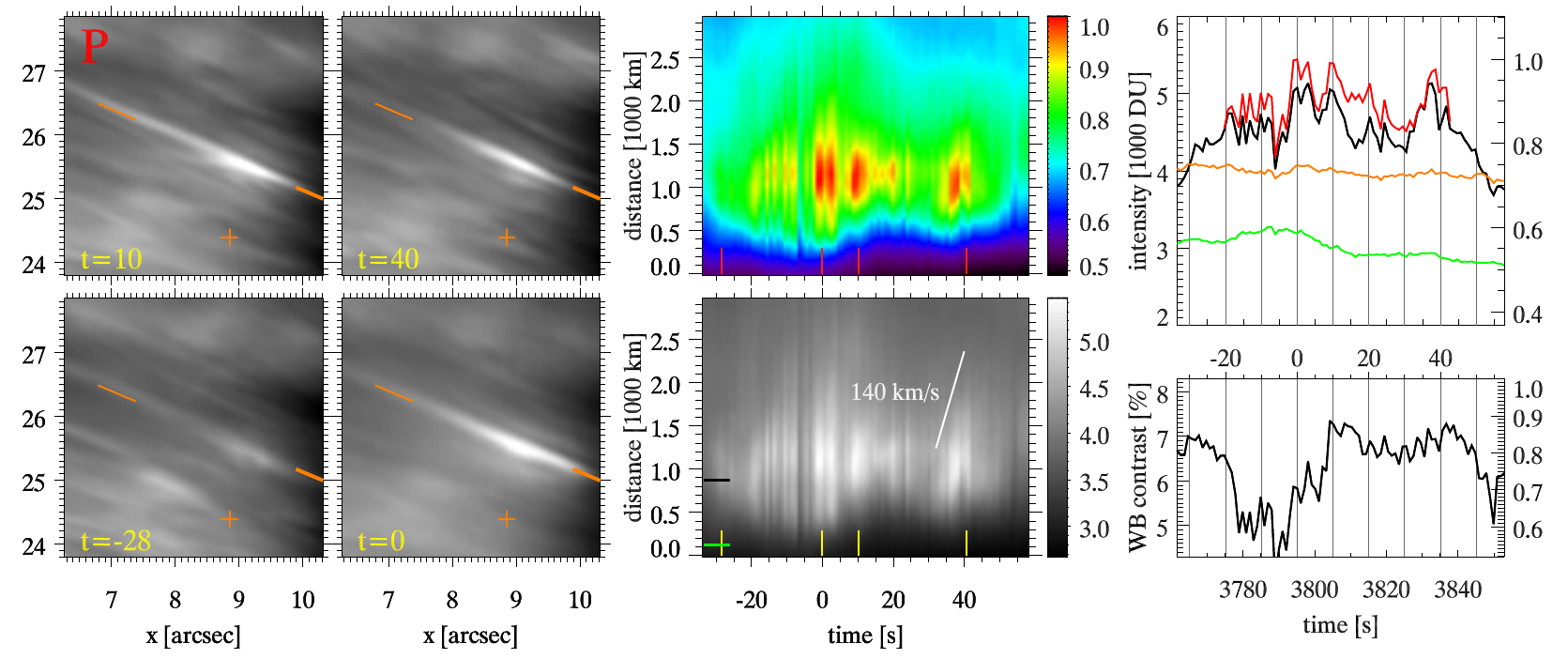}
\caption{\label{fig:app_detailsNOP}%
Details of the evolution of PMJs.
The format of this figure is the same as that of Fig.~\ref{fig:detailsAB}. Animations of this figure are available \url{http://folk.uio.no/rouppe/pmj_highcadence/}.
}
\end{figure*}

\begin{figure*}[!ht]
\includegraphics[width=\textwidth]{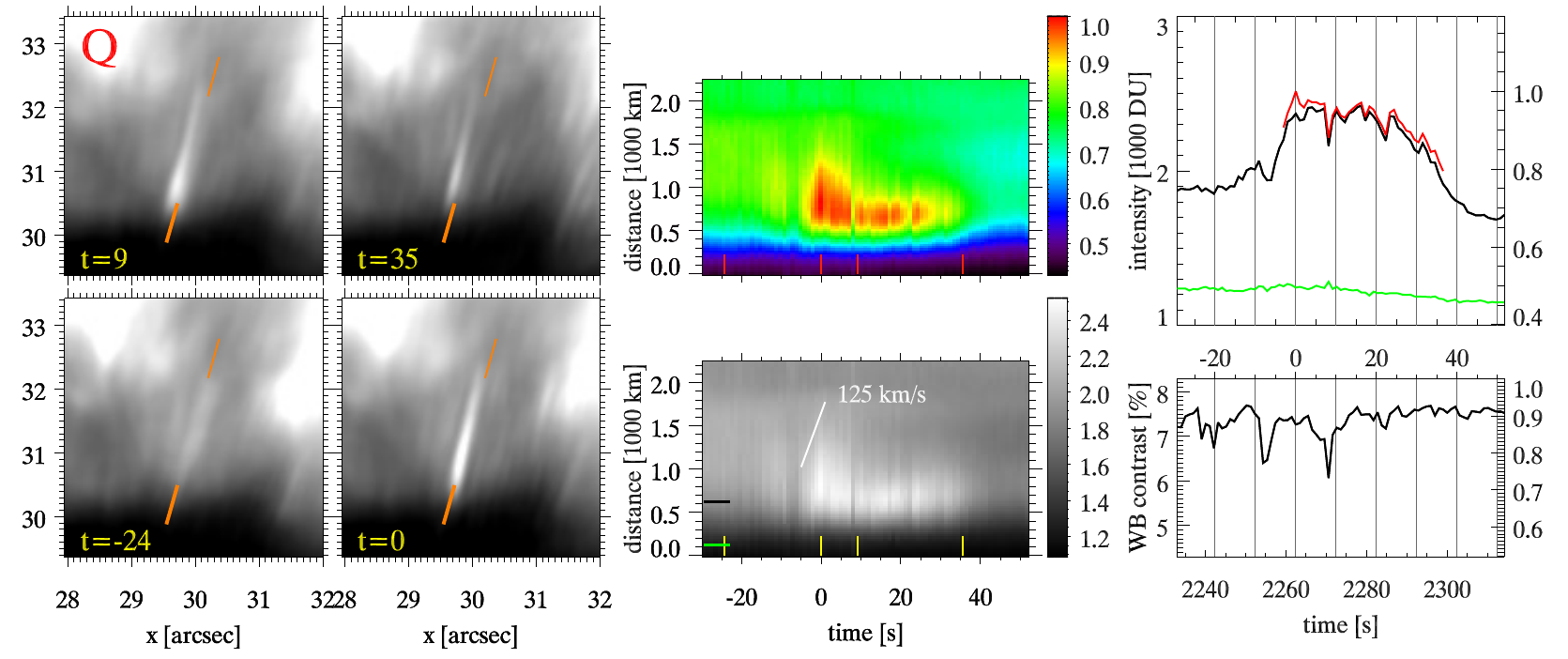}
\caption{\label{fig:app_detailsQ}%
Details of the evolution of a PMJ.
The format of this figure is the same as that of Fig.~\ref{fig:detailsAB}. An animation of this figure is available \url{http://folk.uio.no/rouppe/pmj_highcadence/}.
}
\end{figure*}

\begin{figure*}
\includegraphics[width=\columnwidth]{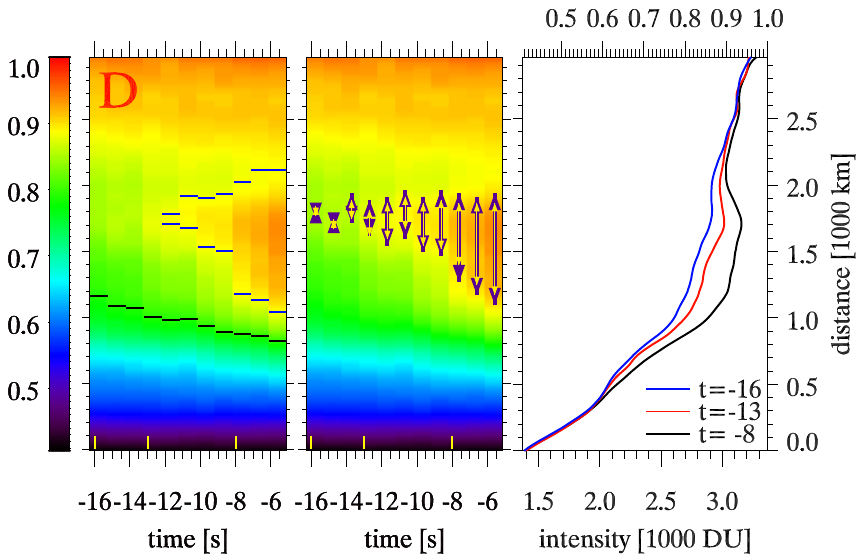}
\includegraphics[width=\columnwidth]{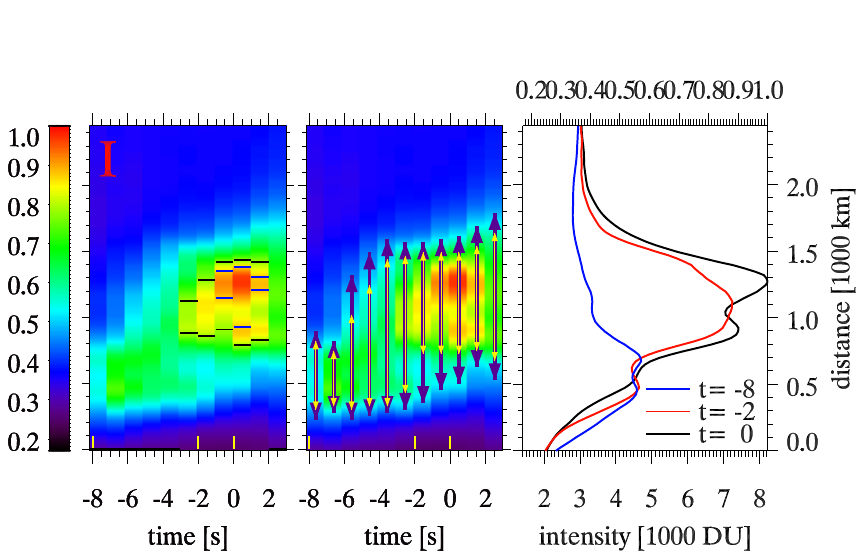} \\
\includegraphics[width=\columnwidth]{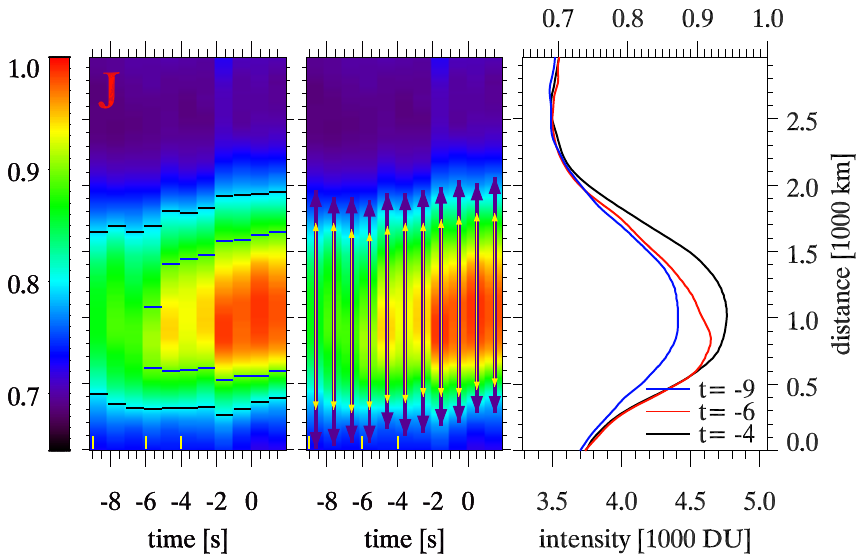}
\includegraphics[width=\columnwidth]{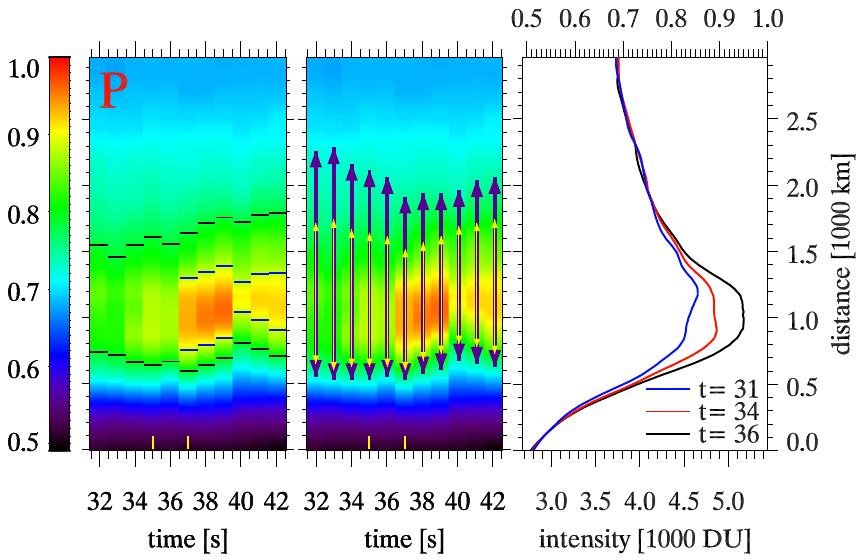} \\
\includegraphics[width=\columnwidth]{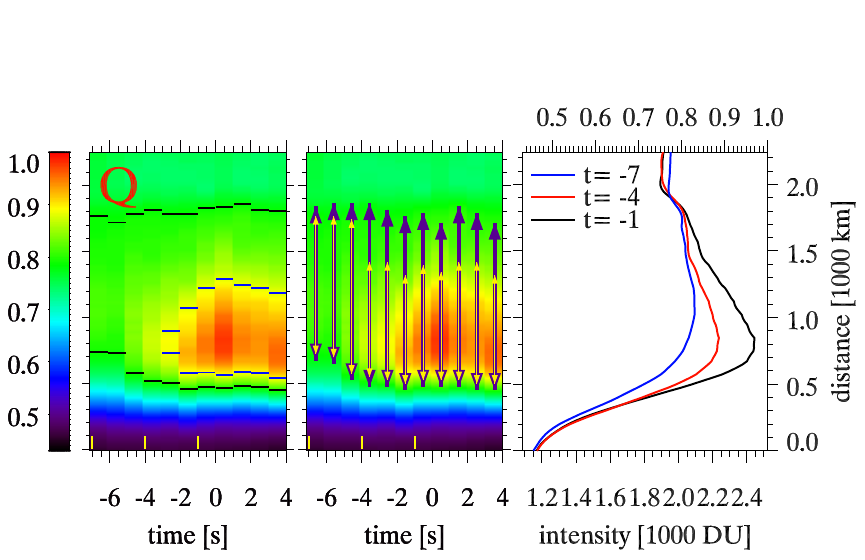}
\caption{\label{fig:app_onset}%
Detailed look at the onset of PMJs. 
The format of this figure is the same as that of Fig.~\ref{fig:onset}.
}
\end{figure*}

\end{appendix}

\end{document}